\documentclass[10pt,aps,twocolumn,nofootinbib,article,letterpaper]{revtex4}\usepackage{amsmath}    
\usepackage{amsfonts} 
\usepackage[utf8]{inputenc}
\usepackage{amssymb}
\usepackage{float}
\usepackage{amsmath}
\usepackage{latexsym}
\usepackage[english]{babel}
\newsavebox\inset
\usepackage{graphicx}
\usepackage{pgfplots}
\usepackage{tikz}
\usepackage{pgfplotstable}
\usepackage{braket}
\definecolor{Mygreen}{rgb}{0.15,0.75,0.45}
\usetikzlibrary{patterns}
\pgfplotsset{compat=newest}
\usepackage{mathtools}
\usepackage{graphics}
\usepackage{lipsum}
\usepackage{cmap}
\usepackage{epsfig}
\usepackage{epstopdf}
\usepackage{color}
\usepackage{commath}
\usepackage[T1]{fontenc}
\usepackage{layout}
\usepackage{listings}
\usepackage{tikz,pgfplots}
\usepackage{bbding}
\usepackage{wasysym}
\def\bibsection{\section*{\refname}}
\DeclareMathOperator{\sech}{sech}
\makeatletter 
\newlength{\oldparindent}
\newcommand{\myindent}{\hspace{\oldparindent}}
\newcommand*{\phoneA}{{\fontspec{code2000.ttf}\symbol{"260E}}}
\newcommand*\Eval[3]{\left.#1\right\rvert_{#2}^{#3}}
\usepgfplotslibrary{fillbetween}
 \usetikzlibrary{arrows}
\selectcolormodel{rgb}
\definecolor{Mygreen}{rgb}{0.15,0.75,0.45}
\usepackage{pict2e}
\usepackage[compact]{titlesec}
\titlespacing{\section}{0pt}{2ex}{2ex}
\renewcommand{\baselinestretch}{1.0}
\newcommand{\ajp}{AJP} 
\begin{document}
\selectcolormodel{cmyk}
\title{Magnetic Stability for the Hatree-Fock Ground State in Two Dimensional Rashba-Gauge Electronic Systems}
\author{\bf{H. Vivas C.}}
\affiliation{Departamento de F\'{i}sica, Universidad
Nacional de Colombia, Sede Manizales, A.A. 127,
Col.}
\email{hvivasc@unal.edu.co}
\date{\today}
\begin{abstract}
The magnetic Hartree Fock ground state stability for a two-dimensional interacting electron system with Rashba-type coupling is studied by implementing the standard many body Green's function formalism. The externally applied electrical field $\mathbf{E}$ enters into the Hamiltonian model through a local gauge-type transformation $\sim E_{j}\mathcal{A}_{j}(\mathbf{k})$, with $\mathcal{A}_{j}(\mathbf{k})$ as the spin gauge vector potential. Phase diagrams associated to the average spin polarization, the Fermi energy, electron density and energy band gap at zero temperature are constructed. The magnetic polarization state as a function of $\mathbf{E}$ is obtained by minimizing the Helmholtz energy functional $\mathcal{F}$ with respect to the average $Z$-spin: $\langle\hat{\sigma}_{Z}\rangle$. We have found that the electric field might reverse the magnetic ground state, unlike the characteristic decreasing associated to the linear $\hat{\boldsymbol{\sigma}}\cdot(\mathbf{k\times E})$ spin-momentum-field coupling.
\end{abstract}
\maketitle
\section{Introduction}
The control of magnetization states on nanoscaled multiferroic devices via externally applied electric field constitutes a technological challenge with multiple and interesting applications \cite{Sp,SG,Heron, Spaldin, Pol, Barnes, Li, Liu, Wang,Kre}. The general consensus for the cross-coupling mechanism of electric/magnetic polarization due to magnetic/electric sources relies on the very first sight upon several phenomena: i) the elastic strain on surfaces or interfaces, ii) the exchange-type Dzyaloshinskii-Moriya interaction and iii) the charge-driven magnetoelectric effects due to charge density accumulation \cite{Vaz, Song}. Underlying Physics shall be understood from the local symmetry properties associated to the \emph{spin} concept.  A spin gauge field vector in this context comes from the space-time induced phase difference between two correlated electronic states, which might be calculated from the first order expansion on the SU(2) generators group, and whose components leads into an effective Rashba \emph{spin-orbit magnetic field} in momentum space $\sim E_{j}\hat{\boldsymbol{\sigma}}\cdot (\mathbf{n}\times\partial_{j}\mathbf{n})$, with $\mathbf{n}=(-k_{y},k_{x},0)/k$ as the unitary vector tangent to the \emph{magnetic texture} on the $XY$ plane, $E_{j}$ as the applied electric field and $\hat{\boldsymbol{\sigma}}$ as the set of Pauli matrices \cite{Tan,Tatara,Na,Nak,Liu2,Tan2}. Attempts for the calculation of the Hartree Fock ground state have been performed by different techniques in terms of the Wigner-Seitz radius $r_{s}\sim 1/\sqrt{n_{s}}$, ($n_{s}$ as the particle density), predicting partial spin polarization due to the Coulomb exchange-driven on semiconducting structures \cite{arg,ital}. In this scenario, we study the magnetic ground state stability in two dimensional and low density interacting electron gas at zero temperature. Electron charge interaction is taken into account via Coulomb contact-type exchange, while the energy band gap is biased by the external electric field for the linear and non-linear (gauge) momentum form. An average out-of-plane spin polarization $\langle\hat{\sigma}_{Z}\rangle$ emerges in a finite range of the Coulomb exchange strength. Analytical results for asymptotic behavior on $\langle\hat{\sigma}_{Z}\rangle$ at zero applied field and negligible kinetic energy are also discussed.
\section{ Hartree Fock Self-Consistent Formulation for Transverse Magnetization}
Single electron dynamics in two-dimensional systems might be described by using a Zeeman-type Hamiltonian $\hat{\mathcal{H}}_{\Sigma}=-\gamma\hat{\boldsymbol{\sigma}}\cdot\mathcal{B}_{\Sigma}$, for $\gamma\mathcal{B}_{\Sigma}$ taken as the effective magnetic field containing the Rashba constant $\alpha$, the electric field intensity $E_{x}$ and the electron momentum $\mathbf{k}$: $\gamma\mathcal{B}_{\Sigma}=(-\alpha k_{y},\alpha k_{x},-\Delta_{\mathbf{k}})$. The last component contains the spin gauge effect for this particular symmetry and the usual linear momentum-field interaction $\hat{\boldsymbol{\sigma}}\cdot(\mathbf{k\times E})$:
\begin{equation}\label{gapp}
\Delta_{\mathbf{k}}=\Delta_{0}+\frac{eE_{x}k_{y}}{k_{0}^{2}}\left(1-\frac{k_{0}^{2}}{k^{2}}\right), 
\end{equation}
with $k_{0}=m^{\star}\alpha/\hbar^{2}$ \cite{Vivas, Landau}. The unitary vector $\mathbf{n}$ is on the same direction of the magnetic field generated by the Rashba interaction $\gamma\mathcal{B}_{R}=(-\alpha k_{y},\alpha k_{x},0)$, i.e., $\mathbf{n}=\mathcal{B}_{R}/\abs{\mathcal{B}_{R}}$ and the spin gauge vector potential points along to the perpendicular direction of $\mathcal{B}_{R}$. The parameter $\Delta_{0}$ represents the band gap size at zero field. In the framework of the Hartree-Fock approximation, the interacting contact (short range) electron-electron Coulomb energy may be written in terms of the $Z$-spin Pauli matrix under the decoupling scheme \cite{Yosida,Fa}: 
\begin{equation}
\mathcal{\hat{H}}_{Z}=-J\langle\hat{\sigma}_{Z}\rangle\hat{\sigma}_{Z}+\frac{J}{2}\langle\hat{\sigma}_{Z}\rangle^{2}\hat{I},
\end{equation}
 where $\hat{I}$ is the $2\times 2$ unit matrix and $\langle\hat{\sigma}_{Z}\rangle$ is taken as the average spin polarization on $Z$ direction. The complete Rashba-Hartree-Fock Hamiltonian reads:
\begin{equation}\label{hHF}
\hat{\mathcal{H}}_{RHF}=\hat{\mathcal{H}}_{K}+\hat{\mathcal{H}}_{Z}+\hat{\mathcal{H}}_{\Sigma},  
\end{equation}
with $\hat{\mathcal{H}}_{K}=(\hbar^{2}k^{2}/2m^{\star})\hat{I}$. The eigenvalues of $\hat{\mathcal{H}}_{RHF}$ are: 
\begin{equation}\label{ref3}
E_{\mathbf{k}\pm}=\frac{\hbar^{2}k^{2}}{2m^{\star}}+\frac{J}{2}\langle\hat{\sigma}_{Z}\rangle^{2}\pm\varepsilon_{\mathbf{k}}^{0},
\end{equation}
with $\varepsilon_{\mathbf{k}}^{0}=(\alpha^{2}k^{2}+\Delta_{J\mathbf{k}}^{2})^{1/2}$ and $\Delta_{J\mathbf{k}}=\Delta_{\mathbf{k}}-J\langle\hat{\sigma}_{Z}\rangle$. Average spin polarization is obtained by calculating the Green's propagator $\mathcal{G}^{0}(k)$ associated to $\mathcal{\hat{H}}_{RHF}$ under the prescription $\langle\hat{\sigma}_{Z}\rangle=(\hbar\beta)^{-1}\sum_{k}\mbox{Tr}\lbrace\hat{\sigma}_{Z}\mathcal{G}^{0}(k)\rbrace$ \cite{Fetter}, with:
\begin{equation}\label{G0}
\mathcal{G}_{ij}^{0}\left(\mathbf{k},i\omega_{n}\right)=\sum_{\sigma=\lbrace\pm\rbrace}\frac{M^{\sigma}_{ij}\left(\mathbf{k}\right)}{i\omega_{n}-\hbar^{-1}(E_{\mathbf{k}\sigma}-\mu)},
\end{equation}
$\omega_{n}=(2n+1)/\hbar\beta$, ($n=0,1,2...$) as the set of fermionic Matsubara frequencies and $\mathbf{M}^{\sigma}\left(\mathbf{k}\right)=\ket{u_{\mathbf{k}\sigma}}\bra{ u_{\mathbf{k}\sigma}}$ corresponds to the matrix of $\hat{\mathcal{H}}_{RHF}$ eigenstates. Normalized eigenvectors are given by $\bra{u_{\mathbf{k}\sigma}}\equiv (-i\sigma e^{i\phi_{\mathbf{k}}}F_{\mathbf{k}\sigma},1)/(1+F^{2}_{\mathbf{k}\sigma})^{1/2}$, while $M^{\sigma}_{ij}\left(\mathbf{k}\right)$ is defined through:
\begin{equation}
M^{\sigma}_{ij}\left(\mathbf{k}\right)=\frac{1}{(1+F^{2}_{\mathbf{k}\sigma})}
\begin{pmatrix}
  F^{2}_{\mathbf{k}\sigma}& i\sigma F_{\mathbf{k}\sigma}e^{-i\phi_{\mathbf{k}}} \\
-i\sigma F_{\mathbf{k}\sigma}e^{i\phi_{\mathbf{k}}}&1\\
 \end{pmatrix},
\end{equation}
with $\alpha kF_{\mathbf{k}\sigma}=\sigma\Delta_{J\mathbf{k}}+(\alpha^{2}k^{2}+\Delta_{J\mathbf{k}}^{2})^{1/2}$ and $\tan{\phi_{\mathbf{k}}}=k_{y}/k_{x}$. By using (\ref{G0}), the average spin polarization takes the self consistent form:
\begin{equation}\label{eq1}
\langle\hat{\sigma}_{Z}\rangle=\sum_{\mathbf{k}\subseteq\mathcal{D}}\frac{\Delta_{J\mathbf{k}}}{\sqrt{\alpha^{2}k^{2}+\Delta_{J\mathbf{k}}^{2}}}\rho_{S}(\mathbf{k}),
\end{equation}
with $\rho_{S}(\mathbf{k})=\sinh{(\beta\varepsilon_{\mathbf{k}}^{0})}/[\cosh{(\beta\mu_{J\mathbf{k}})}+\cosh{(\beta\varepsilon_{\mathbf{k}}^{0})}]$ and $\mu_{J\mathbf{k}}=\mu-(J/2)\langle\hat{\sigma}_{Z}\rangle^{2}-\hbar^{2}k^{2}/2m^{\star}$ as the modified chemical potential. In the zero temperature limit and zero field, the distribution function $\rho_{S}(\mathbf{k})$ evolves towards the step function, with $\rho_{S}(\mathbf{k})=1$ in the domain $\mathcal{D}$ consisting of a circular disk with of radii $\bar{k}_{\mp}=[2(1+\bar{\mu}_{J0}\mp (1+2\bar{\mu}_{J0}+\bar{\Delta}_{J0}^{2})^{1/2})]^{1/2}$ and $\rho_{S}(\mathbf{k})=0$, otherwise. The factors $\Delta_{J0}$, $J$, $\alpha k$ and $\mu_{J0}$ have been normalized (and labeled with an overlying bar) to the reference Rashba energy $\alpha k_{0}$. Also, $\bar{E}_{x}=eE_{x}/\alpha k_{0}^2$.  With $\alpha=4.62\times 10^{-12}$ eV$\cdot$m, which is typical for semiconducting InGaAs/InP asymmetric quantum wells, $k_{0}^{-1}$ falls into the range of $44.63$ nm with $m^{\star}=0.37m_{0}$ and $76.38$ nm for $m^{\star}=0.04m_{0}$ in 2DEG In$_{x}$Ga$_{1-x}$As \cite{QW,Nech}. For tunable Rashba energy on Bi$_{x}$Pb$_{1-x}$/Ag (111) alloys, $\alpha$ reaches the value $3\times 10^{-10}$ eV$\cdot$m, with $m^{\star}=0.3m_{0}$ and $1.2<k_{0}^{-1}<2.5$ nm \cite{Ast}. In the former case, the electrical field reaches a maximum value of $4.6$ V$\cdot$cm$^{-1}$ for $\bar{E}_{x}\approx 0.2$.
\section{Results}
The average spin polarization value $\langle\hat{\sigma}_{Z}\rangle$ (Eq. \ref{eq1}) at zero temperature might also be obtained by minimizing the Helmholtz free energy functional $\delta\bar{\mathcal{F}}/\delta\langle\hat{\sigma}_{Z}\rangle=0$, with $\bar{\mathcal{F}}=\sum_{\mathbf{k}s}(\bar{E}_{\mathbf{k}\sigma}-\bar{\mu})$ integrated under the domain $\mathcal{D}$: $\bar{E}_{\mathbf{k}\sigma}\leq\bar{\mu}$. Figure (\ref{figH}) shows the minimum-shift for the normalized Helmholtz energy $\bar{\mathcal{F}}$ as a function of $\langle\hat{\sigma}_{Z}\rangle$ at zero temperature. Stronger exchange coupling $\bar{J}$ depletes the average magnetization on $Z$ direction. Eye guide dots are shown as reference for the results obtained in Figures (\ref{fig1}) and (\ref{fig2}). Increasing field effects over the local minimum of $\mathcal{F}$ are represented by the continuous lines (a)-(f). Inversion in the sign of $\langle\hat{\sigma}_{Z}\rangle$ as well as its magnitude are possible for some specific values of $\bar{E}_{x}$. Figure (\ref{fig1}) illustrates the phase diagram for the density of particles (the number of particles per unit of area $\mathcal{S}$) at zero field as a function of $\langle\hat{\sigma}_{Z}\rangle$. Electron density is also defined in terms of the Green's propagator as $n_{\bar{\mu}}=(\hbar\beta)^{-1}\sum_{k}\mbox{Tr}\lbrace\mathcal{G}^{0}(k)\rbrace=\sum_{\mathbf{k}\subseteq\mathcal{D}}\rho_{N}(\mathbf{k})$, with the density function number defined by: $\rho_{N}(\mathbf{k})=1+\sinh{(\beta\mu_{J\mathbf{k}})}/[\cosh{(\beta\mu_{J\mathbf{k}})}+\cosh{(\beta\varepsilon_{\mathbf{k}}^{0})}]$. In the gapless limit ($\bar{\Delta}_{0}=0$), no exchange ($\bar{J}=0$), no applied field ($\bar{E}_{x}=0$) and $\bar{\mu}>0$, $n_{\bar{\mu}}=(k_{0}^{2}/\pi)[1+\bar{\mu}-(1+2\bar{\mu})^{1/2}]$, while it decreases as $\bar{\Delta}_{0}$ augments at fixed $\bar{\mu}$ \cite{nn}. In the low-density electron gas approximation, $E_{\mathbf{k}}\approx\varepsilon_{\mathbf{k}}^{0}$, the energy band takes the Dirac cone-like structure, and in case in which the gauge interaction coupling is exclusively considered in Eq. (\ref{gapp}) (i.e., $\Delta_{\mathbf{k}}\approx -eE_{x}k_{y}/k^{2}$), the electron density reads: $n^{G}_{\bar{\mu}}(\bar{E}_{x})=(k_{0}^{2}\bar{\mu}^{2}/\pi^{2})\mathcal{E}(4\bar{E}_{x}^{2}/\bar{\mu}^{4})$. The function $\mathcal{E}(4\bar{E}_{x}^{2}/\bar{\mu}^{4})$ is the complete elliptic integral of the second kind, which is real in the interval $0<4\bar{E}_{x}^{2}/\bar{\mu}^{4}<1$, or $\bar{E}_{x}$ restrained into $\lbrace 0,\bar{\mu}^{2}/2\rbrace$, range in which the Fermi surface remains topologically connected. The electron density lies on the interval $k_{0}^{2}\bar{\mu}^{2}/\pi^{2}<n^{G}_{\bar{\mu}}(\bar{E}_{x})<k_{0}^{2}\bar{\mu}^{2}/2\pi$. An estimation for the density of states (DOS) at zero temperature can be obtained from $n^{G}_{\bar{\mu}}$ under the definition $\mathcal{D}_{\bar{\mu}}(\bar{E}_{x})=\partial n^{G}_{\bar{\mu}}/\partial\bar{\mu}=(2\bar{\mu}/\pi)\mathcal{D}_{F}\mathcal{K}(4\bar{E}_{x}^{2}/\bar{\mu}^{4})$, where $\mathcal{K}(4\bar{E}_{x}^{2}/\bar{\mu}^{4})$ represents the complete elliptic integral of first kind, and $\mathcal{D}_{F}=m^{\star}/\pi\hbar^{2}$ corresponds to the DOS for two dimensional non interacting electron systems. In the limit $\bar{\mu}^{4}>>4\bar{E}_{x}^{2}$ (or $\bar{E}_{x}=0$), DOS behaves in a linear form with $\mathcal{D}_{\bar{\mu}}(0)\approx\bar{\mu}\mathcal{D}_{F}$, while it reveals a strong peak at $\bar{\mu}=(2\bar{E}_{x})^{1/2}$. Under the same set of conditions and with $\Delta_{\mathbf{k}}\approx eE_{x}k_{y}/k_{0}^{2}$, the density of particles is given by $n^{D}_{\bar{\mu}}(\bar{E}_{x})=k_{0}^{2}\bar{\mu}^{2}/2\pi(1+\bar{E}_{x}^2)^{1/2}$. The carrier density $n_{\bar{\mu}}^{G}$ decreases more rapidly than $n_{\bar{\mu}}^{D}$ in the range $0<\bar{E}_{x}\leq\bar{\mu}^{2}/2$. For the full Hamiltonian (\ref{hHF}) at zero field, the solutions of $\langle\hat{\sigma}_{Z}\rangle$ tend towards $\bar{\Delta}_{0}/\bar{J}$ when $n_{\bar{\mu}}>>1$ (in units of $k_{0}^{2}$).
\begin{figure}
\centering
\resizebox{\linewidth}{!}{%
\pgfplotsset{every axis/.append style={font=\large,
extra description/.code={5
}}}

\begin{tikzpicture}
\node at (4.750,6.50){\large{(f)}};
\node at (4.750,4.750){\large{(d)}};
\node at (5.50,5.50){\large{(e)}};
\node at (4.725,2.00){\large{(b)}};
\node at (5.35,1.250){\large{(a)}};
\node at (7.850,5.50){\small{$\bar{J}=1.0$}};
\node at (7.50,3.50){\small{$\bar{J}=0.5$}};
\node at (1.0,2.45){\small{$\bar{J}=0.25$}};
\node at (1.750,0.65){\small{$\bar{J}\approx 0.025$}};
\node at (4.750,3.50){\large{(c)}};
\begin{axis}[scale only axis,
scale=1.0,
domain=0:0.5,xmax=0.5,xmin=-0.5,ymin=-5, ymax=0.0,
xlabel={\Large{$\langle\hat{\sigma}_{Z}\rangle$}},
ylabel={\Large{$\mathcal{\bar{F}}\times 10^{-3}$}},
tick label style={/pgf/number format/fixed,
/pgf/number format/precision=3},
]
\draw[black,thin] (axis cs:0.0,-5.0) -- (axis cs:0.0,0.0);
\draw[thick,black,fill=yellow] (-0.0625, -1.78716) circle[radius=3 pt];
\addplot[black,smooth,thick]coordinates{(-0.5, 0.) (-0.4875, 0.) (-0.475, 0.) (-0.4625, 
  0.) (-0.45, -0.000245888) (-0.4375, -0.00412445) (-0.425, 
-0.0129707) (-0.4125, -0.0271204) (-0.4, -0.0469099) (-0.3875, 
-0.0726635) (-0.375, -0.104676) (-0.3625, -0.143184) (-0.35, 
-0.188335) (-0.3375, -0.240139) (-0.325, -0.298429) (-0.3125, 
-0.362845) (-0.3, -0.432882) (-0.2875, -0.507979) (-0.275, 
-0.587582) (-0.2625, -0.671139) (-0.25, -0.758067) (-0.2375, 
-0.847712) (-0.225, -0.939316) (-0.2125, -1.03201) (-0.2, 
-1.12479) (-0.1875, -1.21656) (-0.175, -1.30609) (-0.1625, 
-1.39207) (-0.15, -1.47311) (-0.1375, -1.54779) (-0.125, 
-1.61467) (-0.1125, -1.67232) (-0.1, -1.71938) (-0.0875, 
-1.75459) (-0.075, -1.77689) (-0.0625, -1.78716) (-0.05, 
-1.78577) (-0.0375, -1.77286) (-0.025, -1.7491) (-0.0125, 
-1.71588) (0., -1.67563) (0.0125, -1.63267) (0.025, -1.59744) 
(0.0375, -1.57278) (0.05, -1.55316) (0.0625, -1.53732) (0.075, 
-1.52527) (0.0875, -1.51896) (0.1, -1.5212) (0.1125, -1.52323) 
(0.125, -1.52092) (0.1375, -1.51219) (0.15, -1.49597) (0.1625, 
-1.47181) (0.175, -1.4397) (0.1875, -1.39998) (0.2, -1.35307) 
(0.2125, -1.29848) (0.225, -1.23705) (0.2375, -1.16997) (0.25, 
-1.09843) (0.2625, -1.02364) (0.275, -0.94674) (0.2875, 
-0.868807) (0.3, -0.790834) (0.3125, -0.713715) (0.325, 
-0.638231) (0.3375, -0.565049) (0.35, -0.494727) (0.3625, 
-0.427725) (0.375, -0.364425) (0.3875, -0.305165) (0.4, 
-0.250281) (0.4125, -0.200147) (0.425, -0.155174) (0.4375, 
-0.115753) (0.45, -0.082156) (0.4625, -0.0544946) (0.475, 
-0.0327241) (0.4875, -0.0166849) (0.5, -0.0061417)};
\draw[thick,black,fill=red] (-0.1,-1.27661) circle[radius=3 pt];
\addplot[red,smooth,thick]coordinates{(-0.5, -0.014185) (-0.4875, -0.0268804) (-0.475, -0.043729) 
(-0.4625, -0.0647193) (-0.45, -0.0897392) (-0.4375, -0.118567) 
(-0.425, -0.150904) (-0.4125, -0.186448) (-0.4, -0.224954) 
(-0.3875, -0.266246) (-0.375, -0.310193) (-0.3625, -0.356686) 
(-0.35, -0.405602) (-0.3375, -0.456792) (-0.325, -0.510062) 
(-0.3125, -0.565161) (-0.3, -0.621771) (-0.2875, -0.6795) (-0.275, 
-0.737876) (-0.2625, -0.796342) (-0.25, -0.854255) (-0.2375, 
-0.910888) (-0.225, -0.965433) (-0.2125, -1.01747) (-0.2, 
-1.06693) (-0.1875, -1.11302) (-0.175, -1.15488) (-0.1625, 
-1.1917) (-0.15, -1.22269) (-0.1375, -1.24717) (-0.125, -1.26454) 
(-0.1125, -1.27441) (-0.1, -1.27661) (-0.0875, -1.27133) (-0.075, 
-1.2592) (-0.0625, -1.24163) (-0.05, -1.22148) (-0.0375, 
-1.20595) (-0.025, -1.1945) (-0.0125, -1.18423)(0., -1.17424) (0.0125, -1.16407) (0.025, -1.15344) (0.0375, -1.14217) (0.05, -1.13016) (0.0625, -1.11735) (0.075, -1.10372) (0.0875, 
-1.08927) (0.1, -1.07405) (0.1125, -1.05814) (0.125, -1.04166) (0.1375, -1.02482) (0.15, 
-1.00797) (0.1625, -0.991851) (0.175, -0.978646) (0.1875, 
-0.965304) (0.2, -0.94909) (0.2125, -0.929057) (0.225, -0.904879) 
(0.2375, -0.876588) (0.25, -0.844448) (0.2625, -0.808876) (0.275, 
-0.77039) (0.2875, -0.729564) (0.3, -0.686998) (0.3125, 
-0.643293) (0.325, -0.599027) (0.3375, -0.554745) (0.35, 
-0.51094) (0.3625, -0.46805) (0.375, -0.426296) (0.3875, 
-0.385467) (0.4, -0.345847) (0.4125, -0.307707) (0.425, 
-0.271259) (0.4375, -0.236659) (0.45, -0.204019) (0.4625, 
-0.173414) (0.475, -0.144899) (0.4875, -0.118515) (0.5, 
-0.0943155)};
\draw[thick,black,fill=Mygreen] (-0.025, -0.761999) circle[radius=3 pt];
\addplot[Mygreen,smooth,thick]coordinates{(-0.5, -0.0727788) (-0.4875, -0.0875193) (-0.475, -0.10373) 
(-0.4625, -0.12141) (-0.45, -0.140544) (-0.4375, -0.161103) 
(-0.425, -0.183043) (-0.4125, -0.206296) (-0.4, -0.230775) 
(-0.3875, -0.256365) (-0.375, -0.282928) (-0.3625, -0.310296) 
(-0.35, -0.338271) (-0.3375, -0.366628) (-0.325, -0.395117) 
(-0.3125, -0.423459) (-0.3, -0.451359) (-0.2875, -0.478506) 
(-0.275, -0.504587) (-0.2625, -0.529299) (-0.25, -0.552367) 
(-0.2375, -0.573571) (-0.225, -0.592792) (-0.2125, -0.610079) 
(-0.2, -0.625837) (-0.1875, -0.64153) (-0.175, -0.657214) 
(-0.1625, -0.672257) (-0.15, -0.686439) (-0.1375, -0.699625) 
(-0.125, -0.711717) (-0.1125, -0.722637) (-0.1, -0.732319) 
(-0.0875, -0.740711) (-0.075, -0.747768) (-0.0625, -0.753452) 
(-0.05, -0.757733) (-0.0375, -0.760588) (-0.025, -0.761999) 
(-0.0125, -0.761955) (0., -0.760451) (0.0125, -0.757488) (0.025, -0.753072) (0.0375, -0.747215) (0.05, 
-0.739935) (0.0625, -0.731255) (0.075, -0.721204) 
(0.0875, -0.709816) (0.1, -0.697131) (0.1125, -0.683193) (0.125, 
-0.668052) (0.1375, -0.651762) (0.15, -0.634382) (0.1625, 
-0.615977) (0.175, -0.596616) (0.1875, -0.57637) (0.2, -0.555317) 
(0.2125, -0.533537) (0.225, -0.511116) (0.2375, -0.488142) (0.25, 
-0.464707) (0.2625, -0.440905) (0.275, -0.416835) (0.2875, 
-0.392597) (0.3, -0.368295) (0.3125, -0.344035) (0.325, 
-0.319924) (0.3375, -0.296072) (0.35, -0.27259) (0.3625, 
-0.249592) (0.375, -0.227188) (0.3875, -0.205493) (0.4, 
-0.184613) (0.4125, -0.164656) (0.425, -0.145719) (0.4375, 
-0.12789) (0.45, -0.111244) (0.4625, -0.0958378) (0.475, 
-0.0817121) (0.4875, -0.0688859) (0.5, -0.0573585)};
\draw[thick,black,fill=blue] (0., -2.40635) circle[radius=3 pt];
\addplot[blue,smooth,thick]coordinates{(-0.5, 0.) (-0.4875, 0.) (-0.475, 0.) (-0.4625, 0.) (-0.45, 
  0.) (-0.4375, 0.) (-0.425, 0.) (-0.4125, 0.) (-0.4, 
  0.) (-0.3875, 
  0.) (-0.375, -0.00224593) (-0.3625, -0.0104667) (-0.35, 
-0.0252737) (-0.3375, -0.0472756) (-0.325, -0.0770988) (-0.3125, 
-0.115369) (-0.3, -0.162683) (-0.2875, -0.219576) (-0.275, 
-0.286472) (-0.2625, -0.36363) (-0.25, -0.451073) (-0.2375, 
-0.548509) (-0.225, -0.655276) (-0.2125, -0.770322) (-0.2, 
-0.892295) (-0.1875, -1.01967) (-0.175, -1.15087) (-0.1625, 
-1.28423) (-0.15, -1.41804) (-0.1375, -1.55049) (-0.125, 
-1.67971) (-0.1125, -1.80377) (-0.1, -1.92073) (-0.0875, 
-2.02867) (-0.075, -2.12578) (-0.0625, -2.21039) (-0.05, 
-2.28102) (-0.0375, -2.3365) (-0.025, -2.376) (-0.0125, -2.39918) 
(0., -2.40635) (0.0125, -2.39884) (0.025, -2.38032) (0.0375, 
-2.36228) (0.05, -2.33805) (0.0625, -2.30545) (0.075, -2.27263) (0.0875, 
-2.24584) (0.1, -2.21217) (0.1125, -2.16812) (0.125, -2.11254) (0.1375, -2.04528) (0.15, -1.96675) (0.1625, 
-1.87781) (0.175, -1.77961) (0.1875, -1.67351) (0.2, -1.56104) 
(0.2125, -1.44381) (0.225, -1.32346) (0.2375, -1.20164) (0.25, 
-1.07993) (0.2625, -0.959806) (0.275, -0.842678) (0.2875, 
-0.729833) (0.3, -0.622482) (0.3125, -0.521787) (0.325, 
-0.428848) (0.3375, -0.344626) (0.35, -0.269811) (0.3625, 
-0.204753) (0.375, -0.149475) (0.3875, -0.103741) (0.4, 
-0.067134) (0.4125, -0.0391207) (0.425, -0.0191053) (0.4375, 
-0.00646853) (0.45, -0.000595298) (0.4625, 0.) (0.475, 
  0.) (0.4875, 0.) (0.5, 0.)};
\addplot[black,smooth,thick]coordinates{(-0.5, 0.) (-0.4875, 0.) (-0.475, 0.) (-0.4625, 0.) (-0.45, 
  0.) (-0.4375, 0.) (-0.425, 0.) (-0.4125, 0.) (-0.4, 
  0.) (-0.3875, 0.) (-0.375, 0.) (-0.3625, 0.) (-0.35, 
  0.) (-0.3375, 0.) (-0.325, 0.) (-0.3125, 0.) (-0.3, 
  0.) (-0.2875, 0.) (-0.275, 0.) (-0.2625, 0.) (-0.25, 
  0.) (-0.2375, 0.) (-0.225, 0.) (-0.2125, 0.) (-0.2, 
  0.) (-0.1875, 
  0.) (-0.175, -0.0186673) (-0.1625, -0.114154) (-0.15, 
-0.281864) (-0.1375, -0.511971) (-0.125, -0.794682) (-0.1125, 
-1.12025) (-0.1, -1.47894) (-0.0875, -1.86107) (-0.075, -2.257) 
(-0.0625, -2.65709) (-0.05, -3.05176) (-0.0375, -3.43146) (-0.025, 
-3.78666) (-0.0125, -4.10786) (0., -4.38559) (0.0125, -4.61042) 
(0.025, -4.77294) (0.0375, -4.86377) (0.05, -4.87356) (0.0625, 
-4.7963) (0.075, -4.63927) (0.0875, -4.41309) (0.1, -4.12843) 
(0.1125, -3.79596) (0.125, -3.42638) (0.1375, -3.03043) (0.15, 
-2.61887) (0.1625, -2.20249) (0.175, -1.7921) (0.1875, -1.39856) 
(0.2, -1.03272) (0.2125, -0.705502) (0.225, -0.427822) (0.2375, 
-0.210643) (0.25, -0.0649514) (0.2625, -0.00176678) (0.275, 
  0.) (0.2875, 0.) (0.3, 0.) (0.3125, 0.) (0.325, 0.) (0.3375, 
  0.) (0.35, 0.) (0.3625, 0.) (0.375, 0.) (0.3875, 0.) (0.4, 
  0.) (0.4125, 0.) (0.425, 0.) (0.4375, 0.) (0.45, 0.) (0.4625, 
  0.) (0.475, 0.) (0.4875, 0.) (0.5, 0.)};
\addplot[black,smooth,thick,densely dashdotted]coordinates{(-0.5, 0.) (-0.4875, 0.) (-0.475, 0.) (-0.4625, 0.) (-0.45, 
  0.) (-0.4375, 0.) (-0.425, 0.) (-0.4125, 0.) (-0.4, 
  0.) (-0.3875, 0.) (-0.375, 0.) (-0.3625, 0.) (-0.35, 
  0.) (-0.3375, -0.00149339) (-0.325, -0.0235661) (-0.3125, 
-0.070583) (-0.3, -0.140706) (-0.2875, -0.232121) (-0.275, 
-0.343035) (-0.2625, -0.471681) (-0.25, -0.616314) (-0.2375, 
-0.775213) (-0.225, -0.94668) (-0.2125, -1.12904) (-0.2, 
-1.32064) (-0.1875, -1.51985) (-0.175, -1.72507) (-0.1625, 
-1.93471) (-0.15, -2.14722) (-0.1375, -2.36106) (-0.125, 
-2.57473) (-0.1125, -2.78672) (-0.1, -2.99558) (-0.0875, 
-3.19988) (-0.075, -3.39817) (-0.0625, -3.58908) (-0.05, 
-3.77123) (-0.0375, -3.94327) (-0.025, -4.10387) (-0.0125, 
-4.25174) (4.51028*10^-17, -4.38559) (0.0125, -4.50417) (0.025, 
-4.60625) (0.0375, -4.69063) (0.05, -4.7561) (0.0625, -4.80152) 
(0.075, -4.82575) (0.0875, -4.82766) (0.1, -4.80616) (0.1125, 
-4.76061) (0.125, -4.69203) (0.1375, -4.60188) (0.15, -4.49166) 
(0.1625, -4.36286) (0.175, -4.21704) (0.1875, -4.05574) (0.2, 
-3.88054) (0.2125, -3.69306) (0.225, -3.49492) (0.2375, -3.28776) 
(0.25, -3.07327) (0.2625, -2.85314) (0.275, -2.62909) (0.2875, 
-2.40287) (0.3, -2.17625) (0.3125, -1.95102) (0.325, -1.72898) 
(0.3375, -1.51199) (0.35, -1.30189) (0.3625, -1.10058) (0.375, 
-0.909957) (0.3875, -0.731958) (0.4, -0.568535) (0.4125, 
-0.421666) (0.425, -0.293353) (0.4375, -0.185619) (0.45, 
-0.100512) (0.4625, -0.0401022) (0.475, -0.00648407) (0.4875, 
  0.) (0.5, 0.)};
\addplot[black,smooth,thick,dashed]coordinates{(-0.5, -0.329864) (-0.4875, -0.40088) (-0.475, -0.477409) 
(-0.4625, -0.559094) (-0.45, -0.645584) (-0.4375, -0.736532) 
(-0.425, -0.831596) (-0.4125, -0.930438) (-0.4, -1.03272) 
(-0.3875, -1.13813) (-0.375, -1.24633) (-0.3625, -1.357) (-0.35, 
-1.46984) (-0.3375, -1.58453) (-0.325, -1.70077) (-0.3125, 
-1.81826) (-0.3, -1.93671) (-0.2875, -2.05583) (-0.275, -2.17532) 
(-0.2625, -2.29493) (-0.25, -2.41435) (-0.2375, -2.53334) (-0.225, 
-2.65162) (-0.2125, -2.76893) (-0.2, -2.88501) (-0.1875, 
-2.99962) (-0.175, -3.11251) (-0.1625, -3.22343) (-0.15, 
-3.33215) (-0.1375, -3.43844) (-0.125, -3.54207) (-0.1125, 
-3.64282) (-0.1, -3.74047) (-0.0875, -3.83481) (-0.075, -3.92563) 
(-0.0625, -4.01272) (-0.05, -4.0959) (-0.0375, -4.17496) (-0.025, 
-4.24972) (-0.0125, -4.31998) (4.51028*10^-17, -4.38559) (0.0125, 
-4.44635) (0.025, -4.50211) (0.0375, -4.55269) (0.05, -4.59793) 
(0.0625, -4.63769) (0.075, -4.6718) (0.0875, -4.70014) (0.1, 
-4.72254) (0.1125, -4.73888) (0.125, -4.74903) (0.1375, -4.75286) 
(0.15, -4.75025) (0.1625, -4.74108) (0.175, -4.72523) (0.1875, 
-4.70261) (0.2, -4.67311) (0.2125, -4.63668) (0.225, -4.59349) 
(0.2375, -4.54376) (0.25, -4.48772) (0.2625, -4.42561) (0.275, 
-4.35764) (0.2875, -4.28408) (0.3, -4.20516) (0.3125, -4.12114) 
(0.325, -4.03226) (0.3375, -3.9388) (0.35, -3.84102) (0.3625, 
-3.73918) (0.375, -3.63356) (0.3875, -3.52444) (0.4, -3.41211) 
(0.4125, -3.29684) (0.425, -3.17894) (0.4375, -3.05871) (0.45, 
-2.93643) (0.4625, -2.81243) (0.475, -2.68702) (0.4875, -2.5605) 
(0.5, -2.43321)};
\addplot[black,smooth,thick,dotted]coordinates{(-0.5, -2.11083) (-0.4875, -2.18144) (-0.475, -2.252) (-0.4625, 
-2.32245) (-0.45, -2.39272) (-0.4375, -2.46276) (-0.425, 
-2.53251) (-0.4125, -2.60192) (-0.4, -2.67093) (-0.3875, 
-2.73948) (-0.375, -2.80753) (-0.3625, -2.87503) (-0.35, 
-2.94191) (-0.3375, -3.00814) (-0.325, -3.07365) (-0.3125, 
-3.1384) (-0.3, -3.20235) (-0.2875, -3.26544) (-0.275, -3.32763) 
(-0.2625, -3.38888) (-0.25, -3.44914) (-0.2375, -3.50836) (-0.225, 
-3.5665) (-0.2125, -3.62353) (-0.2, -3.6794) (-0.1875, -3.73407) 
(-0.175, -3.7875) (-0.1625, -3.83965) (-0.15, -3.89049) (-0.1375, 
-3.93998) (-0.125, -3.98808) (-0.1125, -4.03476) (-0.1, -4.07998) 
(-0.0875, -4.12372) (-0.075, -4.16593) (-0.0625, -4.2066) (-0.05, 
-4.24568) (-0.0375, -4.28314) (-0.025, -4.31896) (-0.0125, 
-4.35312) (4.51028*10^-17, -4.38559) (0.0125, -4.41633) (0.025, 
-4.44532) (0.0375, -4.47255) (0.05, -4.49798) (0.0625, -4.52159) 
(0.075, -4.54336) (0.0875, -4.56328) (0.1, -4.58132) (0.1125, 
-4.59745) (0.125, -4.61167) (0.1375, -4.62396) (0.15, -4.6343) 
(0.1625, -4.64267) (0.175, -4.64906) (0.1875, -4.65346) (0.2, 
-4.65585) (0.2125, -4.65622) (0.225, -4.65456) (0.2375, -4.65086) 
(0.25, -4.64511) (0.2625, -4.6373) (0.275, -4.62742) (0.2875, 
-4.61547) (0.3, -4.60145) (0.3125, -4.58533) (0.325, -4.56713) 
(0.3375, -4.54683) (0.35, -4.52444) (0.3625, -4.49995) (0.375, 
-4.47336) (0.3875, -4.44468) (0.4, -4.4139) (0.4125, -4.38103) 
(0.425, -4.3461) (0.4375, -4.30916) (0.45, -4.27025) (0.4625, 
-4.22941) (0.475, -4.1867) (0.4875, -4.14215) (0.5, -4.0958)};
\addplot[black,smooth,thick,dash dot]coordinates{(-0.5, -4.17333) (-0.4875, -4.18068) (-0.475, -4.18793) (-0.4625, 
-4.19507) (-0.45, -4.20212) (-0.4375, -4.20906) (-0.425, -4.2159) 
(-0.4125, -4.22263) (-0.4, -4.22926) (-0.3875, -4.23579) (-0.375, 
-4.24221) (-0.3625, -4.24853) (-0.35, -4.25475) (-0.3375, 
-4.26086) (-0.325, -4.26687) (-0.3125, -4.27277) (-0.3, -4.27856) 
(-0.2875, -4.28425) (-0.275, -4.28983) (-0.2625, -4.29531) (-0.25, 
-4.30069) (-0.2375, -4.30595) (-0.225, -4.31111) (-0.2125, 
-4.31616) (-0.2, -4.32111) (-0.1875, -4.32595) (-0.175, -4.33068) 
(-0.1625, -4.33531) (-0.15, -4.33982) (-0.1375, -4.34423) (-0.125, 
-4.34854) (-0.1125, -4.35273) (-0.1, -4.35681) (-0.0875, 
-4.36079) (-0.075, -4.36466) (-0.0625, -4.36842) (-0.05, 
-4.37207) (-0.0375, -4.37561) (-0.025, -4.37905) (-0.0125, 
-4.38237) (2.77556*10^-17, -4.38559) (0.0125, -4.38869) (0.025, 
-4.39169) (0.0375, -4.39458) (0.05, -4.39735) (0.0625, -4.40002) 
(0.075, -4.40258) (0.0875, -4.40503) (0.1, -4.40737) (0.1125, 
-4.4096) (0.125, -4.41172) (0.1375, -4.41372) (0.15, -4.41562) 
(0.1625, -4.41741) (0.175, -4.41909) (0.1875, -4.42066) (0.2, 
-4.42212) (0.2125, -4.42346) (0.225, -4.4247) (0.2375, -4.42583) 
(0.25, -4.42685) (0.2625, -4.42775) (0.275, -4.42855) (0.2875, 
-4.42924) (0.3, -4.42981) (0.3125, -4.43028) (0.325, -4.43063) 
(0.3375, -4.43088) (0.35, -4.43101) (0.3625, -4.43104) (0.375, 
-4.43095) (0.3875, -4.43076) (0.4, -4.43045) (0.4125, -4.43003) 
(0.425, -4.42951) (0.4375, -4.42887) (0.45, -4.42813) (0.4625, 
-4.42727) (0.475, -4.42631) (0.4875, -4.42523) (0.5, -4.42405)};
\addplot[thick,smooth,violet]coordinates{(-0.5, 0.) (-0.4875, 0.) (-0.475, 0.) (-0.4625, 0.) (-0.45, 
  0.) (-0.4375, 0.) (-0.425, 0.) (-0.4125, 0.) (-0.4, 
  0.) (-0.3875, 0.) (-0.375, 0.) (-0.3625, 0.) (-0.35, 
  0.) (-0.3375, 0.) (-0.325, 0.) (-0.3125, 0.) (-0.3, 
  0.) (-0.2875, 
  0.) (-0.275, -0.0001024) (-0.2625, -0.00647944) (-0.25, 
-0.0240038) (-0.2375, -0.0545164) (-0.225, -0.0999263) (-0.2125, 
-0.162114) (-0.2, -0.242826) (-0.1875, -0.343573) (-0.175, 
-0.465535) (-0.1625, -0.609495) (-0.15, -0.775777) (-0.1375, 
-0.96422) (-0.125, -1.17417) (-0.1125, -1.40452) (-0.1, -1.65385) 
(-0.0875, -1.92079) (-0.075, -2.20546) (-0.0625, -2.51625) (-0.05, 
-2.83174) (-0.0375, -3.13147) (-0.025, -3.40497) (-0.0125, 
-3.64394) (2.77556*10^-17, -3.84172) (0.0125, -3.99308) (0.025, 
-4.09413) (0.0375, -4.14229) (0.05, -4.13628) (0.0625, -4.07612) 
(0.075, -3.96316) (0.0875, -3.80005) (0.1, -3.59079) (0.1125, 
-3.34072) (0.125, -3.05663) (0.1375, -2.74694) (0.15, -2.42228) 
(0.1625, -2.10159) (0.175, -1.8119) (0.1875, -1.54316) (0.2, 
-1.29425) (0.2125, -1.06627) (0.225, -0.860564) (0.2375, 
-0.678122) (0.25, -0.519434) (0.2625, -0.38443) (0.275, 
-0.272496) (0.2875, -0.182525) (0.3, -0.112993) (0.3125, 
-0.0620582) (0.325, -0.0276813) (0.3375, -0.00774961) (0.35, 
-0.000197458) (0.3625, 0.) (0.375, 0.) (0.3875, 0.) (0.4, 
  0.) (0.4125, 0.) (0.425, 0.) (0.4375, 0.) (0.45, 0.) (0.4625, 
  0.) (0.475, 0.) (0.4875, 0.) (0.5, 0.)};
\linethickness{0.5mm}
\draw[thick,black,fill=Mygreen] (0.2125, -4.65622) circle[radius=3 pt];
\draw[thick,black,fill=gray] (0.1375, -4.75286) circle[radius=3 pt];
\draw[thick,black,fill=blue](0.0875, -4.82766)circle[radius=3 pt];
\draw[thick,black,fill=red] (0.05, -4.87356) circle[radius=3 pt];
\draw[thick,black,fill=yellow](0.3625, -4.43104)circle[radius=3 pt];
\draw[thick,black,fill=violet](0.0375, -4.14229)circle[radius=3pt];
\end{axis}
\end{tikzpicture}
}
\caption{Normalized Helmholtz free energy (in units of $\alpha k_{0}$) as a function of $\langle\hat{\sigma}_{Z}\rangle$ and different electric field intensities for $T=0$, $\bar{J}=2$, $\bar{\mu}=0.5$ and $\bar{\Delta}_{0}=0.1$. (a) $\bar{E}_{x}=0$, (b) $\bar{E}_{x}=0.05$, (c) $\bar{E}_{x}=0.11$, (d) $\bar{E}_{x}=0.15$, (e) $\bar{E}_{x}=0.2$, (f) $\bar{E}_{x}=0.3$. Dotted lines characterize different exchange couplings at zero field. Minimum-shift is observed for increasing values of $\bar{J}>0$. Calculations have been performed under the constraint $\bar{E}_{\mathbf{k}+}<\bar{\mu}$.}\label{figH}
\end{figure}
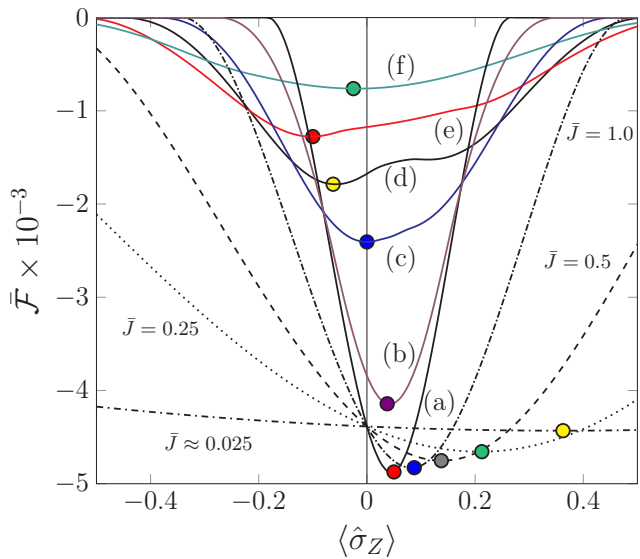
\begin{figure}[h]
  \centering
\resizebox{\linewidth}{!}{%
\pgfplotsset{every axis/.append style={font=\large,
extra description/.code={5
}}}
\begin{tikzpicture}
\node at (3.750,6.50){\large{$\bar{J}=0$}};
\node at (1.250,6.00){\small{$\langle\hat{\sigma}_{Z}\rangle\approx 0.4$}};
\node at (7.50,6.250){\large{$\bar{J}=0.20$}};
\node at (6.50,4.50){\large{$\bar{J}=0.25$}};
\node at (1.250,3.350){\small{$\langle\hat{\sigma}_{Z}\rangle\approx 0.21$}};
\node at (5.50,3.0){\large{$\bar{J}=0.5$}};
\node at (0.850,2.350){\small{$\langle\hat{\sigma}_{Z}\rangle\approx 0.14$}};
\node at (4.50,1.5){\large{$n_{1/2}(\bar{\Delta}_{0}=0.1)$}};
\node at (7.50,1.250){\large{$\bar{J}=2.0$}};
\node at (1.950,0.8750){\small{$\langle\hat{\sigma}_{Z}\rangle\approx 0.045$}};
\begin{axis}[scale only axis,
scale=1,xmode=log,
log ticks with fixed point,
scaled x ticks=real:500
domain=0.001:10,xmax=10,xmin=0.001,ymax=0.501, ymin=0.0,
ylabel={\Large{$\langle\hat{\sigma}_{Z}\rangle$}},
xlabel={\Large{$n_{\bar{\mu}}(k_{0}^{2})$}},
tick label style={/pgf/number format/fixed,
/pgf/number format/precision=3},
]
\draw[black, thick,dashed] (axis cs:0.001,0.141132) -- (axis cs:0.0267463,0.141132);
\draw[black,thick,dashed] (axis cs:0.001,0.0455) -- (axis cs:0.027,0.0455);
\draw[black,thick,dashed] (axis cs:0.027,0.00) -- (axis cs:0.027,0.0455);
\draw[thick,black,fill=yellow] (0.027,0.40) circle[radius=3 pt];
\draw[thick,black,fill=red] (0.027,0.045) circle[radius=3 pt];
\draw[thick,black,fill=Mygreen] (0.027,0.21) circle[radius=3 pt];
\draw[thick,black,fill=gray] (0.027,0.14) circle[radius=3 pt];
\draw[black,thick,dashed] (axis cs:0.001,0.21348) -- (axis cs:0.0265321,0.21348);
\draw[black,thick,dashed] (axis cs:0.0267463,0.0) -- (axis cs:0.0267463,0.141132);
\draw[black,thick,dashed] (axis cs:0.0261827,0.0) -- (axis cs:0.0261827,0.39929);
\draw[black,thick,dashed] (axis cs:0.001,0.39929) -- (axis cs:0.0261827,0.39929);
\addplot[blue,smooth,ultra thick,name path=C]coordinates{(0., 0.)(0.000129566, 0.00501)(0.000265024, 
  0.01002)(0.000407332, 0.01503)(0.000557612, 
  0.02004)(0.000717175, 0.02505)(0.000887567, 
  0.03006)(0.00107061, 0.03507)(0.00126847, 0.04008)(0.0014837, 
  0.04509)(0.00171938, 0.0501)(0.00197919, 0.05511)(0.00226757, 
  0.06012)(0.0025899, 0.06513)(0.00295276, 0.07014)(0.00336421, 
  0.07515)(0.00383421, 0.08016)(0.00437518, 0.08517)(0.00500271,
   0.09018)(0.00573654, 0.09519)(0.00660192, 0.1002)(0.00763149,
   0.10521)(0.00886796, 0.11022)(0.010368, 0.11523)(0.0122077, 
  0.12024)(0.0144912, 0.12525)(0.0173638, 0.13026)(0.0210321, 
  0.13527)(0.0257973, 0.14028)(0.0321116, 0.14529)(0.0406765, 
  0.1503)(0.0526239, 0.15531)(0.069871, 0.16032)(0.0958639, 
  0.16533)(0.137284, 0.17034)(0.208445, 0.17535)(0.344552, 
  0.18036)(0.651914, 0.18537)(1.58372, 0.19038)(7.24703, 
  0.19539)(10.0395,0.19607)};
\addplot[blue,smooth,ultra thick,name path=B]coordinates{(0., 0.)(0.000131874, 0.0051)(0.000269411, 0.0102)(0.00041306, 
0.0153)(0.00056331, 0.0204)(0.000720687, 0.0255)(0.000885762, 
0.0306)(0.00105915, 0.0357)(0.00124153, 0.0408)(0.00143361, 
0.0459)(0.0016362, 0.051)(0.00185015, 0.0561)(0.0020764, 
0.0612)(0.00231595, 0.0663)(0.00256993, 0.0714)(0.00283954, 
0.0765)(0.0031261, 0.0816)(0.00343105, 0.0867)(0.00375599, 
0.0918)(0.00410265, 0.0969)(0.00447293, 0.102)(0.00486893, 
0.1071)(0.00529295, 0.1122)(0.00574753, 0.1173)(0.00623548, 
0.1224)(0.00675988, 0.1275)(0.00732418, 0.1326)(0.00793216, 
0.1377)(0.00858806, 0.1428)(0.00929658, 0.1479)(0.010063, 
0.153)(0.010893, 0.1581)(0.0117933, 0.1632)(0.0127713, 0.1683) 
(0.013835, 0.1734)(0.0149939, 0.1785)(0.0162585, 0.1836) 
(0.0176405, 0.1887)(0.0191536, 0.1938)(0.0208131, 0.1989) 
(0.0226364, 0.204)(0.0246437, 0.2091)(0.026858, 0.2142) 
(0.0293059, 0.2193)(0.0320181, 0.2244)(0.0350304, 0.2295) 
(0.0383843, 0.2346)(0.0421287, 0.2397)(0.0463208, 0.2448) 
(0.0510284, 0.2499)(0.0563321, 0.255)(0.0623282, 0.2601) 
(0.0691324, 0.2652)(0.0768846, 0.2703)(0.0857553, 0.2754) 
(0.0959537, 0.2805)(0.107739, 0.2856)(0.121433, 0.2907) 
(0.137444, 0.2958)(0.15629, 0.3009)(0.178637, 0.306)(0.205357, 
0.3111)(0.237601, 0.3162)(0.276916, 0.3213)(0.325418, 0.3264) 
(0.386062, 0.3315)(0.463064, 0.3366)(0.562609, 0.3417) 
(0.694044, 0.3468)(0.872029, 0.3519)(1.12057, 0.357)(1.48114, 
0.3621)(2.03036, 0.3672)(2.92247, 0.3723)(4.5064, 0.3774) 
(7.71353, 0.3825)(15.7658, 0.3876)(46.6755, 0.3927)(527.246, 
0.3978)};
\addplot[blue,smooth,thick,dashed, name path=F]coordinates{(0., 0.) (0.000129483, 0.00501) (0.000264325, 
  0.01002) (0.00040484, 0.01503) (0.000551363, 
  0.02004) (0.000704246, 0.02505) (0.000863866, 
  0.03006) (0.00103062, 0.03507) (0.00120493, 0.04008) (0.00138725,
   0.04509) (0.00157805, 0.0501) (0.00177784, 0.05511) (0.00198715,
   0.06012) (0.00220657, 0.06513) (0.00243669, 
  0.07014) (0.00267817, 0.07515) (0.00293171, 0.08016) (0.00319803,
   0.08517) (0.00347792, 0.09018) (0.00377222, 
  0.09519) (0.00408183, 0.1002) (0.00440771, 0.10521) (0.00475087, 
  0.11022) (0.00511242, 0.11523) (0.00549352, 0.12024) (0.00589543,
   0.12525) (0.00631949, 0.13026) (0.00676716, 
  0.13527) (0.00723996, 0.14028) (0.	00773958, 0.14529) (0.0082678, 
  0.1503) (0.00882655, 0.15531) (0.00941789, 0.16032) (0.0100441, 
  0.16533) (0.0107075, 0.17034) (0.0114107, 0.17535) (0.0121567, 
  0.18036) (0.0129483, 0.18537) (0.0137889, 0.19038) (0.0146821, 
  0.19539) (0.0156317, 0.2004) (0.0166419, 0.20541) (0.0177174, 
  0.21042) (0.018863, 0.21543) (0.0200843, 0.22044) (0.0213872, 
  0.22545) (0.022778, 0.23046) (0.024264, 0.23547) (0.0258528, 
  0.24048) (0.027553, 0.24549) (0.029374, 0.2505) (0.031326, 
  0.25551) (0.0334203, 0.26052) (0.0356696, 0.26553) (0.0380876, 
  0.27054) (0.0406897, 0.27555) (0.0434929, 0.28056) (0.0465162, 
  0.28557) (0.0497808, 0.29058) (0.0533101, 0.29559) (0.0571305, 
  0.3006) (0.0612716, 0.30561) (0.0657668, 0.31062) (0.0706534, 
  0.31563) (0.0759738, 0.32064) (0.0817761, 0.32565) (0.0881148, 
  0.33066) (0.095052, 0.33567) (0.102659, 0.34068) (0.111017, 
  0.34569) (0.12022, 0.3507) (0.130376, 0.35571) (0.141612, 
  0.36072) (0.154075, 0.36573) (0.167934, 0.37074) (0.183393, 
  0.37575) (0.20069, 0.38076) (0.220108, 0.38577) (0.241985, 
  0.39078) (0.266728, 0.39579) (0.294829, 0.4008) (0.326888, 
  0.40581) (0.363641, 0.41082) (0.406002, 0.41583) (0.455109, 
  0.42084) (0.512401, 0.42585) (0.579716, 0.43086) (0.659422, 
  0.43587) (0.754622, 0.44088) (0.869434, 0.44589) (1.00942, 
  0.4509) (1.18222, 0.45591) (1.39857, 0.46092) (1.67393, 
  0.46593) (2.03115, 0.47094) (2.50518, 0.47595) (3.15158, 
  0.48096) (4.06288, 0.48597) (5.40247, 0.49098) (7.48074, 
  0.49599)(10.09,0.5)};
\addplot[black,smooth,ultra thick,name path=A]coordinates{(0., 0.)(0.000263382, 0.01)(0.00054692, 0.02)(0.000850616, 
  0.03)(0.00117447, 0.04)(0.00151848, 0.05)(0.00188265, 
  0.06)(0.00226697, 0.07)(0.00267145, 0.08)(0.00309609, 
  0.09)(0.00354089, 0.1)(0.00400584, 0.11)(0.00449095, 
  0.12)(0.00499622, 0.13)(0.00552165, 0.14)(0.00606723, 
  0.15)(0.00663297, 0.16)(0.00721887, 0.17)(0.00782492, 
  0.18)(0.00845113, 0.19)(0.0090975, 0.2)(0.00976403, 
  0.21)(0.0104507, 0.22)(0.0111575, 0.23)(0.0118845, 
  0.24)(0.0126317, 0.25)(0.013399, 0.26)(0.0141865, 
  0.27)(0.0149941, 0.28)(0.0158219, 0.29)(0.0166698, 
  0.3)(0.0175379, 0.31)(0.0184262, 0.32)(0.0193346, 
  0.33)(0.0202632, 0.34)(0.0212119, 0.35)(0.0221808, 
  0.36)(0.0231698, 0.37)(0.024179, 0.38)(0.0252084, 
  0.39)(0.0262579, 0.4)(0.0273276, 0.41)(0.0284174, 
  0.42)(0.0295274, 0.43)(0.0306575, 0.44)(0.0318078, 
  0.45)(0.0329783, 0.46)(0.0341689, 0.47)(0.0353796, 
  0.48)(0.0366106, 0.49)(0.0378617, 0.5)};
\addplot[black,smooth,ultra thick,name path=D]coordinates{(0., 0.)(0.0000256907, 0.00101)(0.0000516138, 
  0.00202)(0.0000777993, 0.00303)(0.000104281, 
  0.00404)(0.000131098, 0.00505)(0.000158294, 
  0.00606)(0.000185919, 0.00707)(0.00021403, 
  0.00808)(0.000242692, 0.00909)(0.000271981, 
  0.0101)(0.000301983, 0.01111)(0.000332798, 
  0.01212)(0.000364541, 0.01313)(0.000397346, 
  0.01414)(0.000431371, 0.01515)(0.000466797, 
  0.01616)(0.00050384, 0.01717)(0.000542753, 
  0.01818)(0.000583836, 0.01919)(0.000627446, 
  0.0202)(0.000674009, 0.02121)(0.000724039, 
  0.02222)(0.000778155, 0.02323)(0.000837114, 
  0.02424)(0.000901838, 0.02525)(0.00097347, 
  0.02626)(0.00105343, 0.02727)(0.00114349, 0.02828)(0.00124591,
   0.02929)(0.00136357, 0.0303)(0.0015002, 0.03131)(0.00166069, 
  0.03232)(0.00185151, 0.03333)(0.00208144, 0.03434)(0.00236247,
   0.03535)(0.00271145, 0.03636)(0.00315246, 
  0.03737)(0.00372094, 0.03838)(0.00447058, 0.03939)(0.00548566,
   0.0404)(0.00690444, 0.04141)(0.00896572, 0.04242)(0.0121101, 
  0.04343)(0.0172225, 0.04444)(0.0262924, 0.04545)(0.0445771, 
  0.04646)(0.0899053, 0.04747)(0.257546, 0.04848)(2.37391, 
  0.04949)(6.91655,0.0497)(10.3928,0.049755)};
\addplot fill between[of = A and F, soft clip={domain=0.001:10}, every even segment/.style={blue!10!white}];
\addplot fill between[of = B and C, soft clip={domain=0.001:10}, every even segment/.style={blue!10!white}];
\addplot fill between[of = C and D, soft clip={domain=0.001:10}, every even segment/.style={blue!10!white}];
\addplot fill between[of = F and B, soft clip={domain=0.001:10}, every even segment/.style={blue!10!white}];
\end{axis}
\end{tikzpicture}
}
\caption{Phase correlation for the electron density $n_{\bar{\mu}}$ (in units of $k_{0}^{2}$) and $\langle\hat{\sigma}_{Z}\rangle$ at zero field for $\bar{\Delta}_{0}=0.1$. $n_{1/2}(\bar{\Delta}_{0})$ defines the asymptotic value for the electron density at $\bar{J}=0$ and $\bar{\mu}=1/2$.}\label{fig1}
\end{figure}
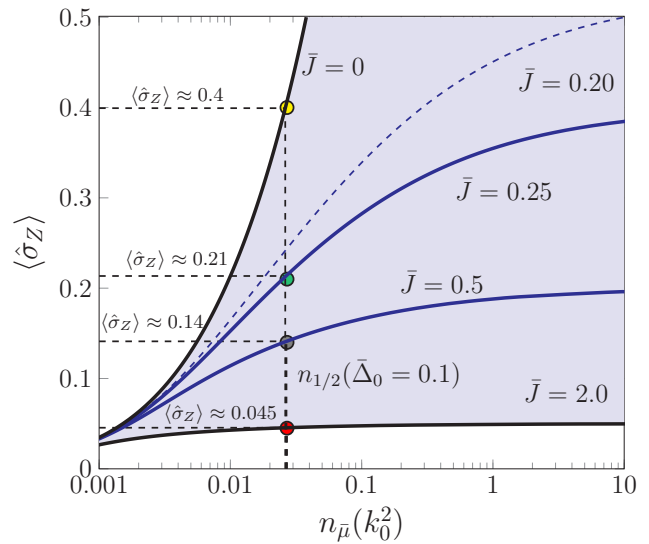
\begin{figure}
\centering
\pgfplotsset{every axis/.append style={
extra description/.code={5
}}}
\begin{tikzpicture}[scale=1,line width=3pt]
\begin{axis}[scale only axis,
scale=0.75,
domain=0:1.0,xmax=1.0, xmin=0.0,ymax=1.5, ymin=0.0,
ytick={0,0.25,0.5,0.75,1,1.25,1.5},
xlabel={\Large{$\bar{\Delta}_{0}$}},
ylabel={\Large{$\bar{\mu}$}}
]
\node at (0.25,0.60) {\textcolor{blue}{$\langle\hat{\sigma}_{Z}\rangle =\frac{1}{2}$}};
\node at (0.675,0.35) {\textcolor{blue}{$\langle\hat{\sigma}_{Z}\rangle =-\frac{1}{2}$}};
\addplot[black,smooth,ultra thick,name path=A]coordinates{(0.0001, 0.0)(0.0102, -0.0000520273)(0.0203, 
-0.000206059)(0.0304, -0.000462096)(0.0405, -0.000820131) 
(0.0506, -0.00128019)(0.0607, -0.00184224)(0.0708, -0.00250632) 
(0.0809, -0.00327242)(0.091, -0.00414051)(0.1011, -0.00511062) 
(0.1112, -0.00618271)(0.1213, -0.00735686)(0.1314, -0.00863298) 
(0.1415, -0.0100111)(0.1516, -0.0114913)(0.1617, -0.0130734) 
(0.1718, -0.0147576)(0.1819, -0.0165438)(0.192, -0.018432) 
(0.2021, -0.0256321)(0.2122, 0.00248466)(0.2223, 
  0.0195427)(0.2324, 0.0360594)(0.2425, 0.0520947)(0.2526, 
  0.0677003)(0.2627, 0.0829207)(0.2728, 0.0977947)(0.2829, 
  0.112356)(0.293, 0.126634)(0.3031, 0.140655)(0.3132, 
  0.154441)(0.3233, 0.168014)(0.3334, 0.18139)(0.3435, 
  0.194587)(0.3536, 0.207617)(0.3637, 0.220496)(0.3738, 
  0.233233)(0.3839, 0.245839)(0.394, 0.258324)(0.4041, 
  0.270696)(0.4142, 0.282964)(0.4243, 0.295133)(0.4344, 
  0.307211)(0.4445, 0.319204)(0.4546, 0.331117)(0.4647, 
  0.342954)(0.4748, 0.354722)(0.4849, 0.366422)(0.495, 
  0.378061)(0.5051, 0.38964)(0.5152, 0.401164)(0.5253, 
  0.412635)(0.5354, 0.424057)(0.5455, 0.435431)(0.5556, 
  0.446762)(0.5657, 0.458049)(0.5758, 0.469297)(0.5859, 
  0.480507)(0.596, 0.49168)(0.6061, 0.502819)(0.6162, 
  0.513925)(0.6263, 0.524999)(0.6364, 0.536044)(0.6465, 
  0.54706)(0.6566, 0.558049)(0.6667, 0.569012)(0.6768, 
  0.57995)(0.6869, 0.590864)(0.697, 0.601755)(0.7071, 
  0.612624)(0.7172, 0.623472)(0.7273, 0.6343)(0.7374, 
  0.645109)(0.7475, 0.655899)(0.7576, 0.666671)(0.7677, 
  0.677426)(0.7778, 0.688164)(0.7879, 0.698887)(0.798, 
  0.709594)(0.8081, 0.720286)(0.8182, 0.730964)(0.8283, 
  0.741628)(0.8384, 0.752279)(0.8485, 0.762918)(0.8586, 
  0.773544)(0.8687, 0.784158)(0.8788, 0.79476)(0.8889, 
  0.805351)(0.899, 0.815932)(0.9091, 0.826502)(0.9192, 
  0.837062)(0.9293, 0.847612)(0.9394, 0.858153)(0.9495, 
  0.868684)(0.9596, 0.879207)(0.9697, 0.889721)(0.9798, 
  0.900227)(0.9899, 0.910724)(1., 0.921214)};
\addplot[black,smooth,ultra thick,name path=B]coordinates{(0.0001,15.1)(0.0102, 11.5592)(0.0203, 3.941)(0.0304, 
  2.23556)(0.0405, 1.54532)(0.0506, 1.18589)(0.0607, 
  0.970826)(0.0708, 0.830492)(0.0809, 0.733461)(0.091, 
  0.663616)(0.1011, 0.611891)(0.1112, 0.572816)(0.1213, 
  0.542906)(0.1314, 0.519841)(0.1415, 0.502016)(0.1516, 
  0.48829)(0.1617, 0.477828)(0.1718, 0.470007)(0.1819, 
  0.464352)(0.192, 0.460494)(0.2021, 0.458145)(0.2122, 
  0.457074)(0.2223, 0.457094)(0.2324, 0.458053)(0.2425, 
  0.459827)(0.2526, 0.462311)(0.2627, 0.46542)(0.2728, 
  0.469078)(0.2829, 0.473225)(0.293, 0.477807)(0.3031, 
  0.482778)(0.3132, 0.488098)(0.3233, 0.493733)(0.3334, 
  0.499652)(0.3435, 0.505831)(0.3536, 0.512244)(0.3637, 
  0.518873)(0.3738, 0.525698)(0.3839, 0.532703)(0.394, 
  0.539875)(0.4041, 0.547199)(0.4142, 0.554664)(0.4243, 
  0.562261)(0.4344, 0.569978)(0.4445, 0.577809)(0.4546, 
  0.585744)(0.4647, 0.593777)(0.4748, 0.601901)(0.4849, 
  0.610111)(0.495, 0.618401)(0.5051, 0.626765)(0.5152, 
  0.635201)(0.5253, 0.643702)(0.5354, 0.652266)(0.5455, 
  0.660889)(0.5556, 0.669567)(0.5657, 0.678298)(0.5758, 
  0.687078)(0.5859, 0.695905)(0.596, 0.704777)(0.6061, 
  0.713691)(0.6162, 0.722645)(0.6263, 0.731637)(0.6364, 
  0.740665)(0.6465, 0.749727)(0.6566, 0.758823)(0.6667, 
  0.76795)(0.6768, 0.777106)(0.6869, 0.786291)(0.697, 
  0.795504)(0.7071, 0.804742)(0.7172, 0.814006)(0.7273, 
  0.823293)(0.7374, 0.832603)(0.7475, 0.841934)(0.7576, 
  0.851287)(0.7677, 0.86066)(0.7778, 0.870053)(0.7879, 
  0.879464)(0.798, 0.888892)(0.8081, 0.898338)(0.8182, 
  0.907801)(0.8283, 0.917279)(0.8384, 0.926773)(0.8485, 
  0.936281)(0.8586, 0.945804)(0.8687, 0.95534)(0.8788, 
  0.96489)(0.8889, 0.974452)(0.899, 0.984027)(0.9091, 
  0.993614)(0.9192, 1.00321)(0.9293, 1.01282)(0.9394, 1.02244)(0.9495, 1.03207)(0.9596, 1.04171)(0.9697, 1.05136)(0.9798, 
  1.06102)(0.9899, 1.07069)(1., 1.08037)};
\addplot[pattern color=yellow!80,pattern=north west lines]fill between[of=A and B];
\end{axis}
\end{tikzpicture}
\caption{Phase space correlation $[\bar{\Delta}_{0},\bar{\mu}]$ for $\abs{\langle\hat{\sigma}_{Z}\rangle}<1/2$ at zero electric field and zero exchange interaction. Inset: $Z$-spin polarization at $\bar{J}=0$ and $\bar{E}_{x}=0$ as a function of the Fermi energy. Solid: $\bar{\Delta}_{0}=0.1$. Dotted: $\bar{\Delta}_{0}=0.25$, Dashed: $\bar{\Delta}_{0}=0.457039$.}\label{P11}
\end{figure}
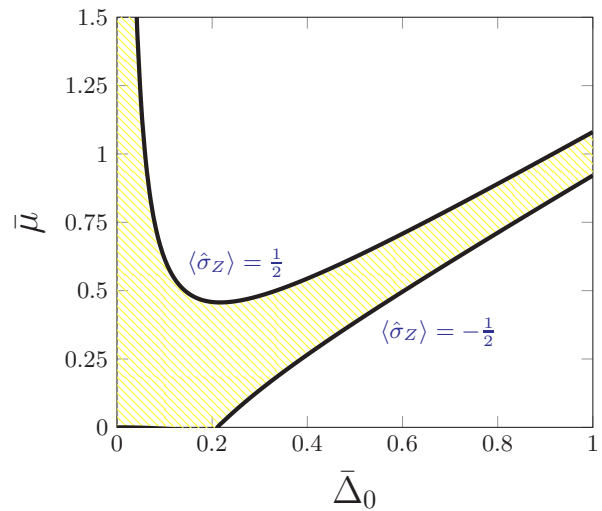
\begin{figure}[h]
  \centering
\resizebox{\linewidth}{!}{%
\pgfplotsset{every axis/.append style={font=\large,
extra description/.code={5
}}}
\begin{tikzpicture}
\begin{axis}[tiny,domain=0.0:1.0,xmax=1.0, xmin=0.0,ymax=0.5, ymin=-0.5,
ticklabel style={font=\small},
xtick={0,0.25,0.5,0.75,1},
ytick={-0.5,-0.25,0.0,0.25,0.5},
ylabel={\large{$\langle\hat{\sigma}_{Z}\rangle$}},
xlabel={\large{$\bar{\mu}$}}
]
\addplot[black,smooth,ultra thick,dotted]coordinates{(0.06, -0.510735) (0.07, -0.481966) (0.08, -0.453435) (0.09, 
-0.425137) (0.1, -0.397065) (0.11, -0.369215) (0.12, -0.341582) 
(0.13, -0.314159) (0.14, -0.286944) (0.15, -0.25993) (0.16, 
-0.233114) (0.17, -0.206491) (0.18, -0.180057) (0.19, -0.153809) 
(0.2, -0.127741) (0.21, -0.101852) (0.22, -0.0761363) (0.23, 
-0.0505914) (0.24, -0.0252137) (0.25, 0.) (0.26, 0.0250528) 
(0.27, 0.0499478) (0.28, 0.074688) (0.29, 0.0992761) (0.3, 
0.123715) (0.31, 0.148007) (0.32, 0.172156) (0.33, 0.196163) 
(0.34, 0.22003) (0.35, 0.243762) (0.36, 0.267359) (0.37, 
0.290824) (0.38, 0.314159) (0.39, 0.337367) (0.4, 0.360449) 
(0.41, 0.383407) (0.42, 0.406244) (0.43, 0.428961) (0.44, 
0.45156) (0.45, 0.474043) (0.46, 0.496412) (0.47, 0.518668)};
\addplot[black,smooth,ultra thick, dashed]coordinates{(0.32, -0.558837) (0.33, -0.516712) (0.34, -0.474812) (0.35, 
-0.433133) (0.36, -0.391672) (0.37, -0.350425) (0.38, -0.30939) 
(0.39, -0.268562) (0.4, -0.227939) (0.41, -0.187518) (0.42, 
-0.147296) (0.43, -0.107269) (0.44, -0.0674356) (0.45, 
-0.0277923) (0.46, 0.0116635) (0.47, 0.0509344) (0.48, 0.0900229) 
(0.49, 0.128932) (0.5, 0.167663) (0.51, 0.206219) (0.52, 
0.244603) (0.53, 0.282817) (0.54, 0.320862) (0.55, 0.358742) 
(0.56, 0.396458) (0.57, 0.434012) (0.58, 0.471407) (0.59, 0.508644)};
\addplot[black,smooth,ultra thick]coordinates{(0., -0.119396)(0.01, -0.106953)(0.02, -0.094631)(0.03, 
-0.0824253)(0.04, -0.0703332)(0.05, -0.0583515)(0.06, 
-0.0464772)(0.07, -0.0347077)(0.08, -0.02304)(0.09, 
-0.0114716)(0.1, -1.34615*10^-15)(0.11, 0.0113772)(0.12, 
0.0226622)(0.13, 0.0338573)(0.14, 0.0449646)(0.15, 0.0559861) 
(0.16, 0.0669238)(0.17, 0.0777796)(0.18, 0.0885552)(0.19, 
0.0992525)(0.2, 0.109873)(0.21, 0.120419)(0.22, 0.130891) 
(0.23, 0.141291)(0.24, 0.15162)(0.25, 0.161881)(0.26, 
0.172073)(0.27, 0.1822)(0.28, 0.192261)(0.29, 0.202258)(0.3, 
0.212193)(0.31, 0.222066)(0.32, 0.231879)(0.33, 0.241632) 
(0.34, 0.251327)(0.35, 0.260965)(0.36, 0.270547)(0.37, 
0.280074)(0.38, 0.289546)(0.39, 0.298965)(0.4, 0.308331) 
(0.41, 0.317646)(0.42, 0.32691)(0.43, 0.336125)(0.44, 0.34529) 
(0.45, 0.354406)(0.46, 0.363475)(0.47, 0.372497)(0.48, 
0.381473)(0.49, 0.390404)(0.5, 0.39929)(0.51, 0.408131)(0.52,
  0.41693)(0.53, 0.425685)(0.54, 0.434398)(0.55, 0.44307) 
(0.56, 0.451701)(0.57, 0.460291)(0.58, 0.468841)(0.59, 
0.477352)(0.6, 0.485825)(0.61, 0.494259)(0.62, 0.502655)};
\end{axis}

\end{tikzpicture}
}
\caption{Inset Figure (3)}
\end{figure}
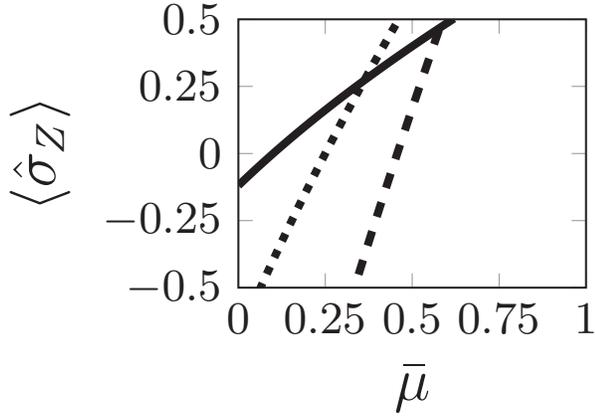
Figure (\ref{P11}) shows the phase space $(\bar{\Delta}_{0},\bar{\mu})$ at $\bar{J}=0$ (shaded area) for physically possible spin range $\abs{\langle\hat{\sigma}_{Z}\rangle}\leq 1/2$. Higher band gap values restrict the allowed Fermi energy (Inset graph). Figure (\ref{fig2}) portrays the main result in this paper: the numerical solutions for $\langle\hat{\sigma}_{Z}\rangle$ as a function of the external applied field $\bar{E}_{x}$ and several coupling strength $\bar{J}$ for conducting electrons dwelling on the positive \emph{helicity} band $E_{\mathbf{k}+}$ \cite{JS}. Comparative results for the usual non-gauge contribution $\Delta_{\mathbf{k}}=\Delta_{0}-\hat{\boldsymbol{\sigma}}\cdot(\mathbf{k\times E})$ are also shown for $0<\bar{J}<2.0$. In this case $\bar{E}_{x}$-scale is reduced in one order of magnitude without inversion in the magnetic orientation \cite{Win}. Inset graph depicts the average spin at zero field and different band gap energies. The change in the curvature of the function $\langle\hat{\sigma}_{Z}\rangle$ is possible for $\bar{\Delta}_{0}>0.33$ and it exhibits a maximum at $\bar{J}\approx 0.27$ for $\bar{\Delta}_{0}=0.4$. At zero field and $\bar{J}=0$, integral (\ref{eq1}) is analytically computable with the result (up to a normalization factor $4\pi^{2}/\mathcal{S}$): $\langle\hat{\sigma}_{Z}\rangle=4\pi k_{0}^{2}\bar{\Delta}_{0}[(1+2\bar{\mu}+\bar{\Delta}_{0}^2)^{1/2}-1-\bar{\Delta}_{0}]$. $\langle\hat{\sigma}_{Z}\rangle$ exhibits a fluctuating behavior in the studied range for the applied electric field with non symmetrical amplitudes and decaying tendency for stronger intensities. Curves resembles the typical \emph{butterfly}-shape associated to the strain intermediate magnetoelastic interaction in ferroelectric/ferromagnetic hybrid structures, in the \emph{direct} polarization cycle \cite{efc,Zh}. $Z$-spin polarization magnitude and its direction are symmetrical under electric field inversion in our model. Parameter $\bar{J}$ weights the repulsive electron-electron contact interaction against the characteristic Rashba coupling $\alpha k_{0}$, indicating that for systems under strong repulsion field, the magnetic ordering is depleted for $\bar{E}_{x}<\bar{E}_{xc}$, with $\bar{E}_{xc}$ as the coercive field value for which the magnetic polarization takes a zero value. For $\bar{E}_{x}>\bar{E}_{xc}$, $\langle\hat{\sigma}_{Z}\rangle$ changes the sign and its amplitude as well as their lowest values are also affected under variations of $\bar{J}$. The coercive field magnitude augments as $\bar{\Delta}_{0}$, and its cutoff value $\bar{E}_{xc0}$ also increases with $\bar{\mu}$, but is independent of $\bar{J}$. A reliable model might be proposed in terms of an exponential-like  behavior: $\bar{\Delta}_{0}=\bar{\mu}(1-\exp{[-\eta(\bar{E}_{xc}^{2}-\bar{E}_{xc0}^{2})]})$, with $\eta$ and $\bar{E}_{xc0}$ as adjusting parameters and $\bar{E}_{xc}\geq\bar{E}_{xc0}$. The relationship $\bar{E}_{xc0}(\bar{\mu})$ is fairly linear in the range of parameters studied.
\begin{figure}
  \centering
\resizebox{\linewidth}{!}{%
\pgfplotsset{every axis/.append style={font=\large,
extra description/.code={5

}}}

\begin{tikzpicture}
\node at (1.5,3.25) {\large{(d)}};
\node at (0.375,4.15) {\large{(c)}};
\node at (1.6,5.65){\large(a)};
\node at (1.5,4.65){\large(b)};
\node at (0.5,2.25){\large(e)};
\node at (7.250,2.5){\large{$\bar{E}_{x}\times 10$}};
\begin{axis}[scale only axis,
scale=1.0,xtick={0,0.05,0.1,0.15,0.2},xticklabels={0,0.05,0.1,0.15,0.2},
domain=0:0.2,xmax=0.2,xmin=0,ymax=0.4, ymin=-0.15,
xlabel={\Large{$\bar{E}_{x}=eE_{x}/\alpha k_{0}^2$}},
ylabel={\Large{$\langle\hat{\sigma}_{Z}\rangle$}},
scaled x ticks=false,
tick label style={/pgf/number format/fixed,
/pgf/number format/precision=2},
]
\addplot[only marks,domain=0:0.2,mark=square*,mark size=2 pt,thick,mark options={solid,fill=yellow}]coordinates{(0., 0.3875) (0.01, 0.375) (0.02, 0.3625) (0.03, 0.3375) (0.04, 
  0.3) (0.05, 0.275) (0.06, 0.25) (0.07, 0.225) (0.08, 
  0.1875) (0.09, 0.1375) (0.1, 0.0625) (0.11, 
  0.) (0.12, -0.05) (0.13, -0.075) (0.14, -0.1) (0.15, -0.1125) 
(0.16, -0.1) (0.17, -0.075) (0.18, -0.05) (0.19, -0.05) (0.2, 
-0.0375)};
\addplot[color=yellow,thick,smooth,domain=0:0.2]{0.399381 - 4.87322*x +  159.238*x^2 - 2895.42*x^3 + 18192.5*x^4 - 36800.8*x^5};
\addplot[only marks,domain=0:0.2,mark=square*,mark size=2 pt,thick,mark options={solid,fill=blue}]coordinates{(0., 0.0875) (0.01, 0.0875) (0.02, 0.075) (0.03, 0.075) (0.04, 
  0.075) (0.05, 0.075) (0.06, 0.075) (0.07, 0.075) (0.08, 
  0.0625) (0.09, 0.0625) (0.1, 0.05) (0.11, 
  0.) (0.12, -0.025) (0.13, -0.05) (0.14, -0.0625) (0.15, 
-0.0875) (0.16, -0.1) (0.17, -0.1125) (0.18, -0.1125) (0.19, 
-0.125) (0.2, -0.125)};
\addplot[color=blue,thick,smooth,domain=0:0.2]{0.0898789 - 0.749242*x - 8.15316*x^2 + 931.817*x^3 - 15464.3*x^4 + 88354.8*x^5 - 167621.0*x^6};
\addplot[only marks,domain=0:0.2,mark=square*,mark size=2 pt,thick,mark options={solid,fill=gray}]coordinates{(0., 0.1375) (0.01, 0.1375) (0.02, 0.1375) (0.03, 0.125) (0.04, 
  0.125) (0.05, 0.1125) (0.06, 0.1125) (0.07, 0.1) (0.08, 
  0.1) (0.09, 0.0875) (0.1, 0.0625) (0.11, 
  0.) (0.12, -0.0375) (0.13, -0.0625) (0.14, -0.0875) (0.15, 
-0.1) (0.16, -0.1125) (0.17, -0.125) (0.18, -0.1125) (0.19, 
-0.0625) (0.2, -0.05)};
\addplot[color=gray,thick,smooth,domain=0:0.2]{0.138147 + 0.711584*x - 88.0235*x^2 + 2495.37*x^3 - 30612.8*x^4 + 156248.0*x^5 - 277971.0*x^6};
\addplot[only marks,domain=0:0.2,mark=square*,mark size=2 pt,thick,mark options={solid,fill=green}]coordinates{(0., 0.2125) (0.01, 0.2) (0.02, 0.2) (0.03, 0.1875) (0.04, 
  0.175) (0.05, 0.1625) (0.06, 0.15) (0.07, 0.1375) (0.08, 
  0.1375) (0.09, 0.1125) (0.1, 0.05) (0.11, 
  0.) (0.12, -0.0375) (0.13, -0.075) (0.14, -0.0875) (0.15, 
-0.1) (0.16, -0.1125) (0.17, -0.1125) (0.18, -0.0625) (0.19, 
-0.05) (0.2, -0.0375)};
\addplot[color=green,thick,smooth,domain=0:0.2]{0.212275 - 0.359933*x - 61.3197*x^2 + 2161.72*x^3 - 30560.4*x^4 +  171413.0*x^5 - 327753.0*x^6};
\addplot[only marks,domain=0:0.2,mark=square*,mark size=2 pt,thick,mark options={solid,fill=red}]coordinates{(0., 0.05) (0.01, 0.05) (0.02, 0.05) (0.03, 0.0375) (0.04, 
  0.0375) (0.05, 0.0375) (0.06, 0.0375) (0.07, 0.0375) (0.08, 
  0.0375) (0.09, 0.0375)(0.11, 
  0.) (0.12, -0.0125) (0.13, -0.0375) (0.14, -0.05) (0.15, 
-0.0625) (0.16, -0.0625) (0.17, -0.075) (0.18, -0.0875) (0.19, 
-0.1) (0.2, -0.1)(0.21,-0.1125)};
\addplot[color=red,thick,smooth,domain=0:0.2]{0.052-0.84*x + 24.73*x^2 - 275*x^3 + 769.7*x^4};
\draw[black,thick,dashed] (axis cs:0.0,0.0) -- (axis cs:0.2,0.0);
\addplot[black,thick,smooth,dashed]coordinates{(0., 0.39929)(0.05, 0.311105) (0.1, 0.176952)(0.15,0.0946182) (0.2, 0.0528523) (0.25, 0.0315122) (0.3,0.0199755) (0.35, 0.0133393) (0.4, 0.00930085) (0.45, 
  0.00672119) (0.5, 0.00500412)};
\end{axis}
\end{tikzpicture}
}
\caption{Numerical solutions for $\langle\hat{\sigma}_{Z}\rangle$, at $\bar{\mu}=0.5$, $\bar{\Delta}_{0}=0.1$. (a) $\bar{J}=0.0$, (b) $\bar{J}=0.25$, (c) $\bar{J}=0.5$, (d) $\bar{J}=1.0$, (e) $\bar{J}=2.0$. Dashed line describes the response for a linear spin-electric field coupling in the form $\sim\hat{\boldsymbol{\sigma}}\cdot(\mathbf{k\times E})$ with $\bar{J}=0$. The horizontal scale might be adjusted by a factor of 10 in this case. Inset: Phase diagram $[\bar{J},\langle\hat{\sigma}_{Z}\rangle$] for different gap magnitudes at zero field, $\langle\hat{\sigma}_{Z}\rangle>0$ and $\bar{\mu}=0.5$. (Dotted) $\bar{\Delta}_{0}=0.1$, (Solid) $\bar{\Delta}_{0}=0.2$, (Squares) $\bar{\Delta}_{0}=0.4$, (Dashed) $\bar{\Delta}_{0}\approx 0.5$.}\label{fig2}
\end{figure}
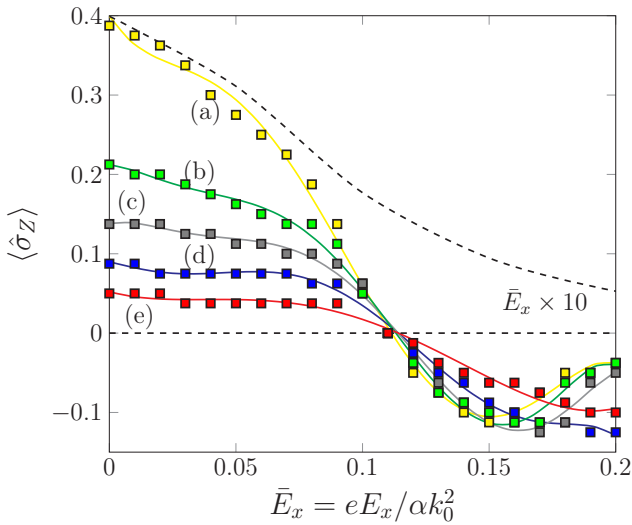
\begin{figure}[h]
  \centering
\resizebox{\linewidth}{!}{%
\pgfplotsset{every axis/.append style={font=\large,
extra description/.code={5
}}}
\begin{tikzpicture}
\begin{axis}[width=5.0 cm,height=4.350 cm,domain=-0.1:2.1,samples=50, xmax=2.1, xmin=-0.1,ymax=0.51,ymin=0.0,
ticklabel style={font=\small},
xtick={0,0.5,1,1.5,2},
ytick={0,0.25,0.5},
ylabel={\large{$\langle\hat{\sigma}_{Z}\rangle$}},
xlabel={\large{$\bar{J}$}},
]
\addplot[black,smooth,thick,dotted]coordinates{(0., 0.39929)(0.1, 0.301106)(0.2, 0.237024)(0.3, 
  0.193931)(0.4, 0.163534)(0.5, 0.141132)(0.6, 0.124011)(0.7, 
  0.110533)(0.8, 0.0996633)(0.9, 0.0907192)(1., 
  0.0832354)(1.1, 0.0768838)(1.2, 0.0714272)(1.3, 
  0.0666898)(1.4, 0.0625391)(1.5, 0.0588727)(1.6, 
  0.0556109)(1.7, 0.0526905)(1.8, 0.0500606)(1.9, 
  0.0476801)(2., 0.0455152)(2.1, 0.043538)(2.2, 0.041725)(2.3,
   0.0400567)(2.4, 0.0385165)(2.5, 0.0370902)(2.6, 
  0.0357656)(2.7, 0.0345322)(2.8, 0.033381)(2.9, 
  0.0323039)(3., 0.0312941)(3.1, 0.0303455)(3.2, 
  0.0294526)(3.3, 0.0286107)(3.4, 0.0278156)(3.5, 
  0.0270634)(3.6, 0.0263508)(3.7, 0.0256748)(3.8, 
  0.0250325)(3.9, 0.0244216)(4., 0.0238398)(4.1, 
  0.023285)(4.2, 0.0227554)(4.3, 0.0222494)(4.4, 
  0.0217654)(4.5, 0.021302)(4.6, 0.0208579)(4.7, 
  0.0204319)(4.8, 0.020023)(4.9, 0.0196301)(5., 0.0192523)};
\addplot[black,smooth,thick]coordinates{(0.1, 0.497)(0.101, 0.496252)(0.201, 0.42152)(0.301, 
  0.359464)(0.401, 0.310383)(0.501, 0.271751)(0.601, 
  0.241012)(0.701, 0.216172)(0.801, 0.195779)(0.901, 
  0.178787)(1.001, 0.164437)(1.101, 0.152173)(1.201, 
  0.141581)(1.301, 0.132346)(1.401, 0.124227)(1.501, 
  0.117036)(1.601, 0.110624)(1.701, 0.104872)(1.801, 
  0.0996843)(1.901, 0.0949822)(2.001, 0.090701)(2.101, 
  0.086787)(2.201, 0.0831951)(2.301, 0.0798873)(2.401, 
  0.0768314)(2.501, 0.0739998)(2.601, 0.0713687)(2.701, 
  0.0689177)(2.801, 0.0666289)(2.901, 0.0644868)(3.001, 
  0.0624778)(3.101, 0.0605899)(3.201, 0.0588124)(3.301, 
  0.057136)(3.401, 0.0555524)(3.501, 0.054054)(3.601, 
  0.0526341)(3.701, 0.0512868)(3.801, 0.0500067)(3.901, 
  0.0487887)(4.001, 0.0476286)(4.101, 0.0465224)(4.201, 
  0.0454662)(4.301, 0.0444569)(4.401, 0.0434914)(4.501, 
  0.0425669)(4.601, 0.0416808)(4.701, 0.0408308)(4.801, 
  0.0400148)(4.901, 0.0392307)};
\addplot[black,smooth, thick,dashed]coordinates{(0., 0.00209277)(0.1, 0.00359114)(0.2, 
  0.0121812)(0.3, 0.227805)(0.4, 0.39299)(0.5, 
0.4502)(0.6,0.45885)(0.7, 0.44614)(0.8, 
0.425)(0.9,0.4012)(1.,0.377556)(1.1, 
0.3552)(1.2, 0.3345)(1.3, 0.3156)(1.4,0.2983)(1.5, 0.282638)(1.6, 0.2684)(1.7, 
0.2553)(1.8, 0.2434)(1.9, 0.23255)(2., 
0.22254)(2.1,0.21332)(2.2, 0.2048)(2.3, 
0.1969)(2.4, 0.189615)(2.5, 0.182812)(2.6, 
  0.176469)(2.7, 0.170543)(2.8, 0.164995)(2.9, 0.15979)(3., 
  0.154899)(3.1, 0.150293)(3.2, 0.145951)(3.3, 0.141849)(3.4, 
  0.137968)(3.5, 0.134293)(3.6, 0.130806)(3.7, 0.127493)(3.8, 
  0.124344)(3.9, 0.121344)(4., 0.118485)(4.1, 0.115757)(4.2, 
  0.11315)(4.3, 0.110658)(4.4, 0.108272)(4.5, 0.105987)(4.6, 
  0.103795)(4.7, 0.101692)(4.8, 0.0996717)(4.9, 0.0977299)(5.,
   0.095862)};
\addplot[only marks,color=blue,mark=square*,mark size=1 pt]coordinates{(0.00, 0.35171)(0.101, 0.428843)(0.201, 0.484438)(0.301, 
  0.50)(0.401, 0.489163)(0.501, 0.461157)(0.601, 
  0.42834)(0.701, 0.395873)(0.801, 0.365833)(0.901, 
  0.33882)(1.001, 0.314807)(1.101, 0.293532)(1.201, 
  0.274668)(1.301, 0.257896)(1.401, 0.242928)(1.501, 
  0.229513)(1.601, 0.217439)(1.701, 0.206525)(1.801, 
  0.19662)(1.901, 0.187596)(2.001, 0.179344)(2.101, 
  0.171772)(2.201, 0.164801)(2.301, 0.158364)(2.401, 
  0.152404)(2.501, 0.14687)(2.601, 0.141718)(2.701, 
  0.136912)(2.801, 0.132417)(2.901, 0.128205)(3.001, 
  0.12425)(3.101, 0.12053)(3.201, 0.117025)(3.301, 
  0.113716)(3.401, 0.110588)(3.501, 0.107626)(3.601, 
  0.104818)(3.701, 0.102152)(3.801, 0.0996172)(3.901, 
  0.0972047)(4.001, 0.0949057)(4.101, 0.0927124)(4.201, 
  0.0906178)(4.301, 0.0886154)(4.401, 0.0866992)(4.501, 
  0.0848639)(4.601, 0.0831043)(4.701, 0.081416)(4.801, 
  0.0797947)(4.901, 0.0782365)};
\draw[black,thin] (axis cs:0.0,0.0) -- (axis cs:0.0,0.5);
\draw[thick,black,fill=yellow] (0.0,0.4) circle[radius=3 pt];
\draw[thick,black,fill=Mygreen] (0.25,0.21) circle[radius=3 pt];
\draw[thick,black,fill=gray] (0.5,0.14) circle[radius=3 pt];
\draw[thick,black,fill=blue] (1.0,0.08) circle[radius=3 pt];
\draw[thick,black,fill=red] (2.0,0.045) circle[radius=3 pt];
\end{axis}
\end{tikzpicture}
}
\caption{Inset Figure (5)}
\end{figure}
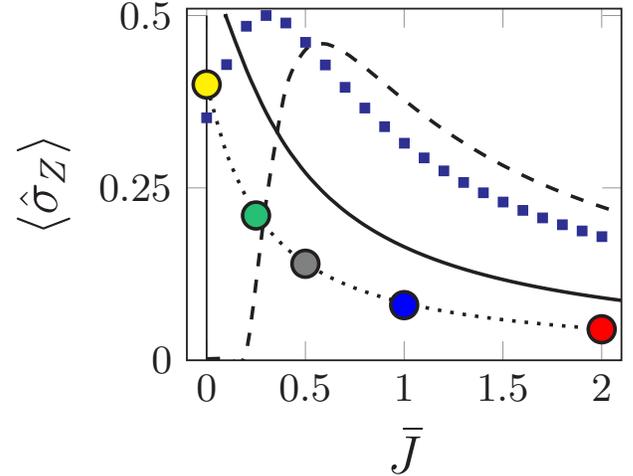
\section{Concluding Remarks}
Possible spin magnetization control under \emph{in-plane} applied electric field has been discussed for two dimensional electron systems with Rashba interaction. The parameter $\langle\hat{\sigma}_{Z}\rangle$ is obtained by implementing a minimization algorithm on the Helmholtz energy functional at zero temperature under topologically constrained Fermi surfaces. The contact electron-electron repulsive effect is considered in the model by using the Hartree-Fock approximation, while the electrical field couples the spin degree of freedom via emerging spin-gauge vector potential. Phase stabilities in the electron density, the average spin polarization and Fermi energy are constructed in the particular case in which $\bar{E}_{x}=0$. We have found in our model that for those systems in which the gauge coupling predominates over the linear spin-momentum-field, it is possible to induce a controlled spin shift inversion for smaller intensities of the external electric field. Non switch orientation in the average magnetization is obtained in the last scenario. Further investigations on additional spin-gauge vector terms might be taken into account, since $\mathcal{A}_{j}(\mathbf{k})$ shall be reconsidered in a higher order expansion scheme when the effective magnetic field $\mathbf{\mathcal{B}}_{\Sigma}$ emerges in the framework of the adiabatic Zeeman-type coupling \cite{Ucr}. Anisotropic energy associated to surface and magnetoelastic strain stimulated by longitudinally applied voltages might also be included into the formalism \cite{aip2,shu}.         


\begin{thebibliography}{99}
\bibitem{Sp} N. A. Spaldin and R. Ramesh, Nature Materials, \textbf{18}, (2019) 203, https://doi.org/10.1038/s41563-018-0275-2
\bibitem{SG} S. T.B. Goennenwein, Europhysics News \textbf{41}, 4 (2010) 17-20, doi:10.1051/epn/2010401
\bibitem{Heron} J. T. Heron \emph{et-al}, Nature \textbf{516}, (2014) 370, doi:10.1038/nature14004 
\bibitem{Spaldin} N.A. Spaldin and M. Fiebig, Science \textbf{309}, (2005) 391, doi: 10.1126/science.1113357
\bibitem{Pol} V. Krivoruchko and A. Savchenko, Acta Physica Polonica A, \textbf{133}, (2018) 463, doi:10.12693/APhysPolA.133.463
\bibitem{Barnes} S. E. Barnes, J. Ieda and S. Maekawa, Sci. Rep. \textbf{4}, (2014) 4105, doi:10.1038/srep04105
\bibitem{Li} C.H. Li, F. Wang, Y. Liu, X.Q. Zhang, Z.H. Cheng and Y. Sun, Phys. Rev. B \textbf{79} 172412 (2009), doi:10.1103/PhysRevB.79.172412
\bibitem{Liu} M. Liu and NX. Sun,  Phil. Trans. R. Soc. A \textbf{372}:
20120439 (2014), doi:10.1098/rsta.2012.0439
\bibitem{Wang} K. F. Wang. J.-M. Liu and Z. F. Ren, Advances in Physics \textbf{58}(4), 321 (2009), doi:10.1080/00018730902920554
\bibitem{Kre}J. Krempask\'y \emph{et-al}, Phys. Rev. X \textbf{8}, (2018) 021067, doi:10.1103/PhysRevX.8.021067
\bibitem{Vaz} C A F Vaz, J. Phys.: Condens. Matter \textbf{24}, (2012) 333201, stacks.iop.org/JPhysCM/24/333201
\bibitem{Song} C. Song, B. Cui, F. Li, X. Zhou and F. Pan, Progress in Materials Science \textbf{87}, (2017) 33, https://doi.org/10.1016/j.pmatsci.2017.02.002
\bibitem{Tan} S.G. Tan \emph{et-al}, Sci. Rep. \textbf{5}, (2015) 18409, doi:10.1038/srep18409
\bibitem{Tatara}  G. Tatara, Physica E: \emph{Low-dimensional
Systems and Nanostructures} (2018), doi:10.1016/j.physe.2018.05.011.
\bibitem{Na} N. Nagaosa, X. Z. Yu and Y. Tokura, Phil. Trans. R. Soc. A (2012) {\bf 370}, 5806–5819, doi:10.1098/rsta.2011.0405
\bibitem{Nak} N. Nakabayashi and G, Tatara,  New J. Phys. \textbf{16}, (2014) 015016, doi:10.1088/1367-2630/16/1/015016
\bibitem{Liu2} W. E. Liu, S. Chesi, D. Webb, U. Z\"{u}licke, R. Winkler, R. Joynt and D. Culcer, Phys. Rev. B \textbf{96}, (2017) 235425, doi:10.1103/PhysRevB.96.235425
\bibitem{Tan2} S. G. Tan, M. B. A. Jalil and T. Fujita, Annals of Physics, \textbf{325}, Issue \textbf{8} (2010) 1537, doi.org/10.1016/j.aop.2010.04.007 
\bibitem{arg} L. O. Juri and P. I. Tamborenea, Phys. Rev. B \textbf{77}, (2008) 233310, doi:10.1103/PhysRevB.77.233310
\bibitem{ital} A. Ambrosetti, F. Pederiva, E. Lipparino and S. Gandolfi, Phys. Rev. B \textbf{80}, (2009) 125306, doi:10.1103/PhysRevB.80.125306
\bibitem{Vivas} H. Vivas C., J. Magn. Magn. Mater. \textbf{449} (2018) 40, https://doi.org/10.1016/j.jmmm.2017.10.004
\bibitem{Landau} Equation  (\ref{gapp}) resembles the solution for the electrostatic potential of a conducting \emph{uncharged} infinite cylinder of radius $R$ placed in a uniform external field $\mathbf{E}$, under $\mathbf{r}\rightarrow\mathbf{k}$, $(\mathbf{E\cdot r})\rightarrow(\mathbf{E\times k})_{Z}$ and $R\rightarrow k_{0}$; see for example: L. D. Landau, E. M. Lifshitz and L. P. Pitaevskii, \emph{Electrodynamics of Continuos Media}, Vol. 8, 2nd ed. Elsevier (2006).  
\bibitem{Yosida} K. Yosida, \emph{Theory of Magnetism}, Springer, Tokyo (1996).
\bibitem{Fa} P. Fazekas, \emph{Series in Modern Condensed Matter Physics -Vol. 5: Lectures Notes on Electron Correlation and Magnetism}, World Scientific Publishing Co., Singapure (1999).
\bibitem{Fetter} A. Fetter, J. D. Walecka, \emph{Quantum Theory of Many Particle Systems}, Dover Publications, New York (2003).
\bibitem{QW} F. Herzog \emph{et al}, New J. Phys. \textbf{19} (2017) 103012, https://doi.org/10.1088/1367-2630/aa833d
\bibitem{Nech} I. A. Nech and E. V. Chulkov, Physics of Solid State, \textbf{51}, N$_{0}.$ 9, (2009) 1772, doi:10.1134/S1063783409090029
\bibitem{Ast} C.R. Ast \emph{et al}, Phys. Rev. B. \textbf{77}, 081407(R) (2008). doi:10.1103/PhysRevB.77.081407
\bibitem{nn} Particle density for the maximum average magnetization $\langle\hat{\sigma}_{Z}\rangle=1/2$ at zero field and $\bar{J}=0$ is given by $n_{\bar{\mu}}(\bar\Delta_{0})=(1+16\pi\bar{\Delta}_{0}^{2})/(128\pi^{3}\bar{\Delta}_{0}^2)$, $\bar{\Delta}_{0}\neq 0$.
\bibitem{JS} J. Sinova \emph{et al}, Phys. Rev. Lett. \textbf{92}, 126603 (2004). https://doi.org/10.1103/PhysRevLett.92.126603
\bibitem{Win} R. Winkler, \emph{Spin-Orbit Coupling Effects in Two-Dimensional Electron and Hole Systems}, Spring-Verlag, Berlin (2003).
\bibitem{efc} S. Gepr\"{a}gs, A. Brandlmaier, M. Opel, R. Gross and S. T. B. Goennenwein, Appl. Phys. Lett. \textbf{96} (2010) 142509, https://doi.org/10.1063/1.3377923
\bibitem{Zh} H. Zhen \emph{et-al}, Science \textbf{303} (2004) 661, doi:10.1126/science.1094207
\bibitem{Ucr} K. Yu. Bliokh and Yu. P. Bliokh, Annals  Phys, \textbf{319} (2005) 13, doi:10.1016/j.aop.2005.03.001
\bibitem{aip2} Y. T. Yang, \emph{et al}, AIP Advances \textbf{7}, 055833 (2017), http://dx.doi.org/10.1063/1.4978588
\bibitem{shu} L. Shu, Z. Li, J. Ma \emph{et-al}, Appl. Phys. Lett. \textbf{100}, 022405 (2012), doi:10.1063/1.3675868 
\end{thebibliography}
\end{document}